\documentclass{emulateapj}

\newcommand{\Code}{\texttt{SEDONA}}
\newcommand{\kms}{\ensuremath{\mathrm{km~s}^{-1}}}
\newcommand{\Nifs}{\ensuremath{^{56}\mathrm{Ni}}}
\newcommand{\Cofs}{\ensuremath{^{56}\mathrm{Co}}}
\newcommand{\Fefs}{\ensuremath{^{56}\mathrm{Fe}}}
\newcommand{\Feff}{\ensuremath{^{54}\mathrm{Fe}}}
\newcommand{\msun}{\ensuremath{M_\odot}}
\newcommand{\texp}{\ensuremath{t_{\mathrm{exp}}}}

\newcommand{\recthreetwo}{\ensuremath{3\rightarrow 2}}
\newcommand{\rectwoone}{\ensuremath{2\rightarrow 1}}
\newcommand{\reconezero}{\ensuremath{1\rightarrow 0}}
\newcommand{\Mch}{\ensuremath{M_{\mathrm{ch}}}}
\newcommand{\alphaexp}{\ensuremath{\alpha_{\mathrm{exp}}}}
\newcommand{\Tto}{\ensuremath{T_{21}}}
\newcommand{\ani}{\ensuremath{a_\mathrm{Ni}}}
\newcommand{\aime}{\ensuremath{a_\mathrm{IME}}}
\newcommand{\Xni}{\ensuremath{X_\mathrm{Ni}}}
\newcommand{\Xime}{\ensuremath{X_\mathrm{IME}}}
\newcommand{\Xca}{\ensuremath{X_\mathrm{ca}}}
\newcommand{\afe}{\ensuremath{a_\mathrm{Fe}}}
\newcommand{\Mni}{\ensuremath{M_\mathrm{Ni}}}
\newcommand{\Mfe}{\ensuremath{M_\mathrm{Fe}}}
\newcommand{\Mime}{\ensuremath{M_\mathrm{IME}}}

\shortauthors{Kasen, D.}
\shorttitle{Near-Infrared Supernova Lightcurves}

\begin{document}

\title{Secondary Maximum in the Near-Infrared Lightcurves of Type~Ia
Supernovae}

\author{Daniel Kasen\altaffilmark{1,2}\email{kasen@pha.jhu.edu}} 
\altaffiltext{1}{Allan C. Davis Fellow, Department of Physics and Astronomy, Johns Hopkins University,
Baltimore, MD 21218}
\altaffiltext{2}{Space Telescope Science Institute, Baltimore, MD 21218}

\begin{abstract} 
We undertake a theoretical study of the near-infrared (NIR)
lightcurves of Type~Ia supernovae (SNe~Ia).  In these bands, the
lightcurves are distinguished by a secondary maximum occurring roughly
20 to 30 days after the initial one.  Using time-dependent multi-group
radiative transfer calculations, we calculate the UBVRIJHK-band
lightcurves of model SN~Ia ejecta structures.  Our synthetic NIR
lightcurves show distinct secondary maxima, and provide favorable fits
to observed SNe~Ia.  We offer a detailed explanation of the origin of
the NIR secondary maximum, which is shown to relate directly to the
ionization evolution of iron group elements in the ejecta.  This
understanding provides immediate intuition into the dependence of the
NIR lightcurves on the physical properties of the ejecta, and in
particular explains why brighter supernovae have a later and more
prominent secondary maximum.  We demonstrate the dependence of the NIR
lightcurves on the mass of \Nifs, the degree of \Nifs\ mixing, the
mass of electron caputre elements, the progenitor metallicity, and the
abundance of intermediate mass elements (especially calcium).  The
secondary maximum is shown to be a valuable diagnostic of these
important physical parameters.  The models further confirm that SNe~Ia
should be excellent standard candles in the NIR, with a dispersion
$\la 0.2$~mag even when the physical properties of the ejecta are
varied widely.  This study emphasizes the consummate value of NIR
observations in probing the structure of SNe~Ia and in furthering
their cosmological utility.
\end{abstract}


\section{Introduction}

The extensive monitoring of Type~Ia supernovae (SNe~Ia) has typically
focused on optical band observations, with only occasional ventures
into the infrared \citep{Kirshner_1973,Elias_1981,Elias_1985}.  The
situation is now beginning to change, and near-infrared (NIR)
observations over the two or three months following explosion are
becoming more common, at least for the more nearby events.  About a
dozen SNe~Ia have published, well observed JHK-band lightcurves
\citep[e.g.,][]{Meikle_IR2000,Hernandez_98bu,Valentini_03E,
Candia_00cx, Kris_2001AJ,Kris_01el,Kris_2004AJ2,Kris_2004AJ1}. NIR
spectra have been obtained for several objects as well
\citep{Meikle_94D, Bowers_86G,Rudy_00cx, Hoeflich_99by, Marion_IR}.
Future SN~Ia surveys promise to gather NIR lightcurves
for a statistically interesting sample of events.  With the broadening
of our wavelength horizons comes the potential to probe the nature of
SN~Ia explosions in new ways.

Observational studies of SNe~Ia in the NIR have often emphasized their
cosmological utility \citep{Kris_IRHubble}.  Normal SNe~Ia are found
to be excellent \emph{standard} (as opposed to calibrated) candles in
the NIR, with an intrinsic dispersion of less than 0.2~mag in the J,H,
and K bands \citep{Elias_1985,Meikle_IR2000}.  Moreover, dust
extinction is smaller by a factor $\sim 5$ in the NIR as compared to
the V band, largely eliminating uncertainties in the reddening
corrections. The primary challenge for the cosmology studies, naturally,
is that SNe~Ia are much fainter and harder to observe in the NIR.

The NIR lightcurves of SNe~Ia possess a morphology distinct from those
at optical wavelengths.  The I through K-band lightcurves are
distinguished by the presence of a secondary maximum, occurring
roughly 20 to 30 days after the initial one.  A corresponding
``shoulder'' can often be seen also in the R and V-band lightcurves.
When the light from all bands is suitably integrated, the bolometric
lightcurves sometimes show an inflection at these times as well
\citep{Contardo_bol}.

In the I-band, the properties of the secondary maximum are found to
correlate with the lightcurve decline rate (and hence peak luminosity)
of the supernova, being more prominent and occurring later in the
broader/brighter SNe~Ia \citep{Hamuy_96Templates, Nobili_Iband}.  Very
subluminous objects may lack a secondary maximum entirely.  Similar
trends appear in the J, H, and K band lightcurves, although there are
interesting exceptions to the rule \citep{Kris_2001AJ,Candia_00cx}.

Occasional theoretical studies have touched upon the NIR lightcurves
of SNe~Ia.  \cite{Hoeflich_DD} computed I-band and NIR lightcurves for
several delayed-detonation models, many of which displayed secondary
maxima.  The authors explained the double-peaked behavior as a
``temperature-radius'' effect, in which the expansion of the
photosphere compensates for the declining temperature in the ejecta.
\cite{Wheeler_IR} and \cite{Hoeflich_99by} have further demonstrated
the value of NIR spectra in diagnosing the physical conditions in the
ejecta. \cite{Pinto-Eastman_II} described the NIR secondary maximum as
the release of pre-existing, trapped radiative energy due to a sudden
decrease in the flux mean opacity.  They placed special emphasis on
the role played by stable iron group elements at the ejecta center in
this process. Most significantly, \cite{Pinto-Eastman_II} recognized
the importance of the ionization state of the ejecta, in particular
the increased NIR emissivity of singly as opposed to double ionized
iron group elements. This last idea is confirmed and extended in the
models studied here.

Despite the theoretical insights, the origin of the NIR secondary
maximum and its dependence on the SN ejecta properties have remained
in many ways, and to many people, obscure.  Moreover, theoretical
models of SNe~Ia have often had difficulty in fitting the distinctive
double-peaked I-band lightcurves
\citep[e.g.,][]{Hoeflich_Khokhlov_LC,Pinto-Eastman_II,Blinnikov_2004}
whereas very few NIR lightcurve calculations have been attempted
\citep[c.f.][]{Hoeflich_DD}.  This paper focuses exclusively on
theoretical models of the far-red and NIR lightcurves of SNe~Ia.  We
calculate the synthetic lightcurves of parameterized ejecta
configuration using the time-dependent multi-dimensional radiative
transfer code \Code\ (\S\ref{Sec:Technique}).  Our model lightcurves
show distinct secondary maxima and provide reasonable fits to the
observed R, I, J and (to a lesser extent) H and K band lightcurves of
SNe~Ia (\S\ref{Sec:Fiducial}).

Close examination of our model calculations provides a clear-cut
explanation for the NIR secondary maximum in SNe~Ia
(\S\ref{Sec:Explain}).  The double-peaked behavior is related directly
to the ionization evolution of iron group elements in the SN
ejecta. In particular, the NIR emissivity of iron/cobalt increases
sharply at a temperature $T \approx 7000$~K marking the transition
between the singly and doubly ionized states.  The eventual cooling of
the iron-rich layers of ejecta to this temperature is thus accompanied
by a sudden increase in emission at far red and NIR wavelengths.

Understanding the origin of the secondary maximum provides immediate
intuition into the dependence of the NIR lightcurves on the ejecta
properties, and allows for theoretical explanations for the
empirically established correlations (\S\ref{Sec:Depend}).  In
particular we study the effect of variations in the mass of \Nifs, the
degree of mixing of \Nifs, the mass of electron capture elements, the
progenitor metallicity, and the abundance of intermediate mass
elements (especially calcium).  All effects are confirmed with
illustrative theoretical models.  The NIR secondary maximum is shown
to be a valuable diagnostic of these important physical parameters.
The models further demonstrate that SNe~Ia are indeed excellent
standard candles in the NIR, even when the physical properties of the
ejecta are varied widely (\S\ref{Sec:Dispersion}).

\section{The Near-IR Emission of SNe~Ia}
\label{Sec:Emission}

\begin{figure}[t]
\begin{center}
\includegraphics[width=8.5cm,clip=true]{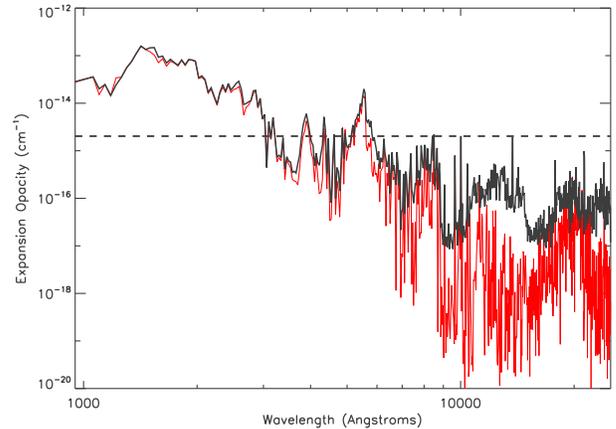}
\caption{Opacity as a function of wavelength of a parcel of
cobalt/iron rich ejecta gas at density $\rho =
10^{-14}~\mathrm{cm}^{-3}$, temperature $T = 15000$~K and time since
explosion \texp = 20~days.  The composition is 75\% \Cofs, 24\% iron
and 1\% \Nifs.  LTE level populations are assumed. The thick black
line shows the bound-bound expansion opacity
(Equation~\ref{Eq:Exp_Opacity}) calculated using the Kurucz CD1 atomic
linelist ($\sim 42$ million lines), while the thin red line is that
calculated using the Kurucz CD1 linelist ($\sim 500,000$ lines).  Note
the large differences in the near-IR.  The horizontal dashed line
marks the electron scattering opacity level.
\label{Fig:Opacity}}
\end{center}
\end{figure}

Before turning to the lightcurve models, we consider the source of NIR
emission in SN~Ia atmospheres.  The luminosity of a SN~Ia is powered
entirely by the decay of radioactive isotopes synthesized in the
explosion, in particular \Nifs\ in the decay chain $\Nifs \rightarrow\
\Cofs \rightarrow\ \Fefs$.  The energetic products of the decay
(primarily gamma-rays) heat the SN ejecta, which radiate thermally.
The bulk of the radiation occurs initially at ultraviolet (UV) and
blue wavelengths.  Direct emission at NIR wavelengths (from bound-free
and free-free sources) is typically of minor significance.

A portion of the blue/UV radiation pervading the SN ejecta is degraded
to red and NIR wavelengths by interaction with lines. The primary
redistribution mechanism is fluorescence \citep{Hoeflich_94D,
Pinto-Eastman_II} whereby a UV photon excites a high-energy atomic
transition, then de-excites via a cascade of longer wavelength
transitions \citep[the reverse process is also possible and occurs as
well;][]{Mazzali_MC}.  The wavelength redistribution of radiation in
atomic lines is responsible for virtually all of the NIR emission
observed from SNe~Ia.  The high velocity gradient in the ejecta
Doppler broadens the radiation from each individual transition, and
the radiation from literally millions of lines blurs together to form
a pseudo-continuum emission in the NIR.

We can calculate, in an approximate way, the NIR opacity and
emissivity due to the large complex of lines.  A convenient way to
capture the average effect is through an expansion opacity formalism
\citep{Karp,Pinto-Eastman_Spectra}, in which individual lines are
summed together to form a pseudo-continuum opacity
\begin{equation}
\alphaexp(\lambda) =
\frac{1}{c \texp}\frac{\lambda}{\Delta\lambda}
\sum_i 
(1 - e^{-\tau_i}),
\label{Eq:Exp_Opacity}
\end{equation}
where \texp\ is the time since explosion.  The sum runs over all lines
in the wavelength bin of size $\Delta \lambda$ at wavelength
$\lambda$.  The quantity $\tau_i$ is the line Sobolev optical depth
\citep{Mihalas_SA}.

In Figure~\ref{Fig:Opacity}, we show the expansion opacity calculated
for a parcel of iron/cobalt rich ejecta gas at density $\rho =
10^{-13}~\mathrm{cm}^{-3}$, temperature $T = 15000$~K and time \texp =
20~days, typical of the inner layers of SNe~Ia.  We use the Kurucz CD1
atomic linelist containing $\sim$ 42 million lines, and compute the
atomic level populations assuming local thermodynamic equilibrium
(LTE). The iron group expansion opacity increases sharply towards the
blue, due to the much larger number of iron group lines occurring at
shorter wavelengths.

In Figure~\ref{Fig:Opacity}, we also compare the expansion opacity
computed using the much smaller (but commonly applied) Kurucz CD~23
linelist, containing only $\sim 500,000$~lines.  Although the two
lists give similar results in the blue, order of magnitude
differences appear in the NIR.  This emphasizes the large number
($\sim 5$ million) of relatively weak lines that may contribute to the
NIR emission of SNe~Ia.

Assuming LTE, the monochromatic emissivity $\eta_\lambda$ is given by the
opacity \alphaexp\ times the blackbody function $B_\lambda(T)$.
Because we are ultimately interested in the total NIR emission over a
broadband wavelength region, we define an average emissivity per gram
(units ergs/s/g/\AA) over the broadband filter $F$ by convolving the
monochromatic emissivity with the appropriate filter transmission
function $\phi_F(\lambda)$
\begin{equation}
\bar{\eta}_F(T)  = \frac{1}{\rho}
\frac{\int_0^\infty d\lambda~\phi_F(\lambda) B(T,\lambda) \alphaexp(T,\lambda)}
{\int_0^\infty d\lambda~\phi_F(\lambda)}.
\label{Eq:Exp_Emission}
\end{equation}
Although this expression formally represents a thermal emission
source, under certain conditions it well approximates the operative
line fluorescence process as well.  In the limit that the time-scales
for radiative transitions are short compared other time-scales of
interest, the radiation field in the lines approaches the
thermodynamic limit embodied by the Planck function.  These are the
conditions that drive the line source function and level populations
close to that of LTE, and should hold approximately at least for the
complex iron group species in the deeper layers of ejecta
\citep{Baron_NLTE}.  Given the dominance of the radiative rates, the
temperature $T$ appearing in Equation~\ref{Eq:Exp_Emission} should be
considered more representative of the radiation field, rather than the
electron gas.  However, the weak coupling through bound-free and
free-free processes may bring the gas and radiation into equilibrium
as well \citep{Pinto-Eastman_II}.

\begin{figure}[t]
\begin{center}
\includegraphics[width=8.5cm,clip=true]{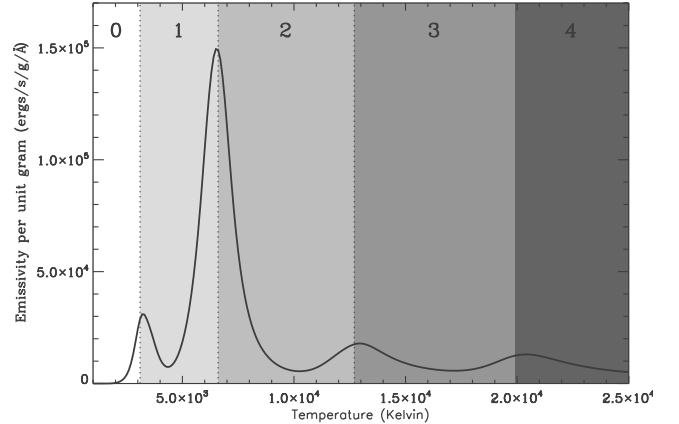}
\caption{Mean I-band emissivity (as a function of temperature) for the
iron/cobalt gas discussed in Figure~\ref{Fig:Opacity} at density $\rho
= 10^{-14}~\mathrm{cm}^{-3}$ and time since explosion \texp = 40~days.
Calculations use Equation~\ref{Eq:Exp_Emission}, the Kurucz CD1 atomic
linelist, and the \cite{Bessell_1990} I-band filter profile. The shading
illustrates the ionization fraction of the gas, which changes very
sharply with temperature at the edges marked in the figure.  A strong
peak in emissivity is seen at the transition between the singly and
doubly ionized states (\rectwoone) occurring at $T \approx 7000$~K.
\label{Fig:Fe_Emis}}
\end{center}
\end{figure}

Note that, even assuming LTE, the wavelength dependence of the
emissivity is not a smooth blackbody function, but depends upon the
the number of lines with substantial opacity in a given part of the
spectrum.  As a result, the NIR emissivity shows a very interesting
non-linear dependence upon temperature. In Figure~\ref{Fig:Fe_Emis}, we
see that the mean I-band emissivity ($\bar{\eta}_I$) of
iron/cobalt-rich gas peaks sharply at temperatures where the gas is
near an ionization edge.  The most prominent peak is at a temperature
$\Tto \approx 7000$~K, marking the transition between the singly and
doubly ionized states (hereafter termed the \rectwoone\ ionization
edge or front).

The interesting dependence of the NIR emissivity on temperature has a
simple explanation in terms of the atomic physics.  NIR radiation
arises from transitions between relatively closely spaced atomic
energy levels.  The atomic levels are generally more closely spaced at
high excitation energy (i.e. high atomic number $n$).  Thus, for a
given ionization state, the NIR emission increases with temperature as
the excited levels become increasingly populated.  When the
temperature becomes hot enough to ionize the species, however, the NIR
emissivity drops rapidly.  The ionized species has a higher overall
energy scale (being more tightly bound) and hence its excited levels
are not highly populated.  If the temperature is increased further,
this behavior is repeated until the next ionization stage is reached.
We emphasize that this emissivity dependence has nothing to do with
the energy released from recombination itself, which makes a
negligible contribution to the overall energy budget.

\begin{figure}
\begin{center}
\includegraphics[width=8.5cm,clip=true]{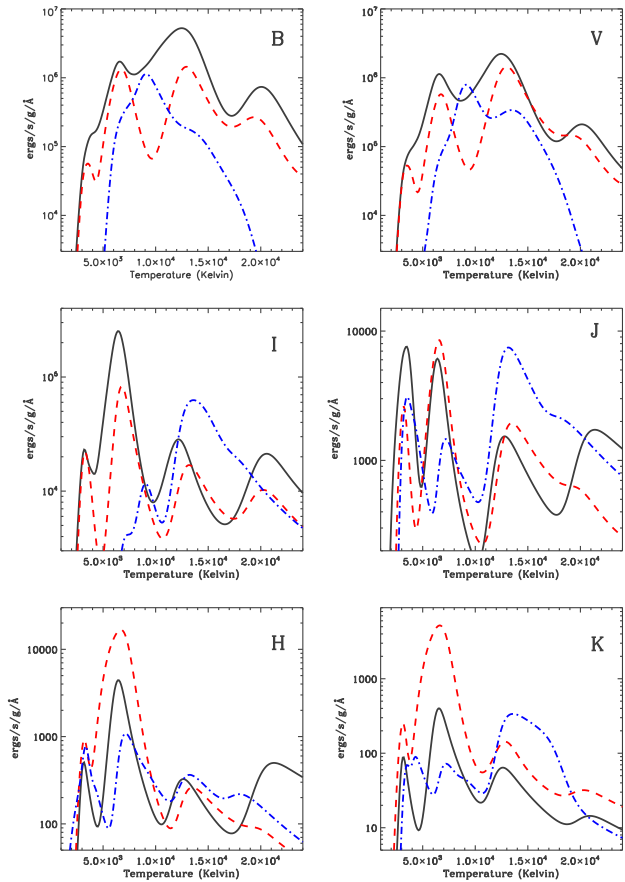}
\caption{A more complete plot of the mean NIR emissivity demonstrated
in Figure~\ref{Fig:Fe_Emis}, this time shown for each of the BVIJHK
bands and for three compositions: pure iron (solid black), pure cobalt
(dashed red) and a mix of 70\% silicon and 30\% sulfur (dot-dashed
blue).  The density is $\rho = 10^{-14}~\mathrm{cm}^{-3}$ and the time
since explosion \texp = 40~days.
\label{Fig:Comp_Emis}}
\end{center}
\end{figure}

For completeness, Figure~\ref{Fig:Comp_Emis} shows the emissivity for
several wavelength bands and for compositions of pure iron, pure
cobalt, and a mixture of silicon and sulfur.  The emissivity of the
iron-group species is qualitatively similar in all NIR bands. A peak
at $\Tto = 7000$~K occurs also in the optical B and V bands, although
it is much less prominent than in the NIR.  The relative size of the
emissivity peaks depends upon the details of atomic structure, and
thus varies somewhat between iron and cobalt and in the different
wavelength regions. In general, the iron-group emissivity exceeds that
of intermediate mass elements such as silicon and sulfur, except at
very high temperatures.  This is because the complex iron group
valence electron structure leads to a extremely large number of line
transitions.

The emissivity behavior shown in Figure~\ref{Fig:Fe_Emis} turns out to
be the key to understanding the NIR secondary maximum in SNe~Ia.  When
the iron rich layers of SN ejecta begin cooling to $\Tto \approx 7000$
(which occurs roughly 35 to 40 days after the explosion) the gas
becomes very effective in redistributing the blue/UV radiation to
longer wavelengths.  This leads to the rebrightening of the NIR
lightcurves.  During these phases, one expects the NIR emission to be
dominated by the relatively thin shell of material located near the
ionization front at \Tto.  As the ejecta progressively cools, this
shell recedes deeper into the ejecta, following the inward propagation
of the \rectwoone\ recombination.

The secondary maximum in the NIR lightcurves is thus identified with the onset
and recession of the \rectwoone\ ionization front into iron-rich ejecta.
In the next few sections we delineate the process in detail, and show
how it serves as an interesting probe of the SN ejecta structure.

\section{Theoretical SNe~Ia Lightcurve Models}
\label{Sec:Technique}

A short time after the explosion of a SN~Ia ($\sim 1$ minute)
hydrodynamic and nucleosynthetic processes abate and the ejected
material reaches a phase of nearly free expansion.  Thereafter, the
essential theoretical challenge becomes to simulate the production and
diffusion of photons through the hot and optically thick ejecta --
i.e., the radiative transfer problem.  Here we calculate broadband
model lightcurves by solving the transfer equation using the
multi-dimensional time-dependent Monte Carlo code \Code\
\citep{Kasen_MC}.

\subsection{Model Ejecta Structures}

To demonstrate the basic dependencies of the NIR lightcurves, we
calculate synthetic \Code\ lightcurves for simple, hand-constructed
SN~Ia ejecta models.  In the free-expansion phase, the velocity of the
ejecta is homologous and everywhere proportional to radius: $r =
v\texp$ where $\texp$ is the time since explosion.  Given the
self-similar nature of the flow, velocity is used as the spatial
coordinate in the simulation.  

Polarization observations of SNe~Ia demonstrate that the asymmetry of
the ejecta is typically not large \citep{Wang_01el}.  We therefore
only consider spherically symmetric ejecta structures in this paper.
Deviations from sphericity will not effect the basic radiative
transfer effects identified here, though they will lead to some
orientation dependence of the model lightcurves.

In 1-dimension (1-D) hydrodynamic explosion models of SNe~Ia such as
W7 \citep{Nomoto_w7}, the final ejecta density structure is well
characterized by an exponential.  We therefore take the density
profile of our ejecta models to be
\begin{equation}
\rho(v,t) = \frac{\Mch}{8\pi v_e^3 t^3} \exp(-v/v_e),
\end{equation}
where \Mch\ is the Chandrasekhar mass and the velocity scale $v_e =
2750$~\kms\ is determined by setting the kinetic energy $KE = 6 \Mch
v_e^2$ equal to the energy released in burning 1.25~\msun\ of the
original carbon/oxygen white dwarf (viz. $1.6\times10^{51}$~ergs).

In analogy to well-known 1-D explosion models such as W7, we tailor a
four-zone stratified compositional structure.  For our fiducial model
these zones are, from the center out: (1) 0.05~\msun\ of stable iron
peak elements (specifically $^{54}$Fe and $^{58}$Ni, typical of
electron capture); (2) 0.6~\msun\ of radioactive \Nifs; (3) 0.6~\msun\
of intermediate mass elements (IME) specifically, Si, S, Ar and Ca, in
proportion to their solar abundances (4) 0.15~\msun\ of unburned
carbon-oxygen with solar metallicity.

To smoothly connect these zones, we use a Gaussian function that
allows us to control the degree of mixing.  For example, mixing
between the \Nifs\ and IME layers is assumed to occur over a mass
range defined by the free parameter \ani.  Above the interface between
the two zones (mass coordinate $m_n = \Mfe + \Mni - \ani/2$) the
nickel abundance is taken to fall off as
\begin{equation}
\Xni = \exp(-(M - m_{n})^2/\ani^2),
\end{equation}
while the IME abundance is $\Xime = 1 - \Xni$.  Similar mixing
parameters are defined for the iron and \Nifs\ interface (\afe) and
the IME and unburned interface (\aime).  In our fiducial model we use
\ani = 0.1, \afe = 0.05, and \aime = 0.1.  In fact, only \ani\ is of
real consequence to the NIR lightcurves (\S\ref{Sec:Mixing}).

The rough compositional structure described, while only suggestive of
detailed explosion models, is quite adequate for our purposes.  The
NIR lightcurves are not sensitive to the detailed trace abundances,
with one important exception -- as we discuss in \S\ref{Sec:Calcium},
small amounts of calcium can potentially affect the I-band, due to
the overwhelming strength of the Ca~II IR triplet feature.
In all other bands, the overriding factor is the amount and
distribution of iron-group elements.

\subsection{Technical Considerations of the Radiative Transfer}

Given a homologously expanding SN ejecta structure, the time-dependent
\Code\ code calculates high resolution synthetic spectra at each day,
from day one to several months after the explosion.  Broadband
lightcurves are then constructed by convolving the spectrum at each
time with the appropriate filter transmission functions.  We use the
filter profiles of \cite{Bessell_1990} for the optical bands and those
of \cite{Persson_IR} for the JHK bands.

\Code\ includes a detailed treatment of gamma-ray transfer to
determine the instantaneous energy deposition rate from radioactive
\Nifs\ and \Cofs\ decay.  Radiative heating and cooling rates are
evaluated from Monte Carlo estimators, and the temperature structure
of the ejecta determined by iterating the model to thermal
equilibrium.  The present calculations use 120 radial zones, 100 time
points, and 5000 wavelength groups.  Resolution tests confirm the
adequacy of this gridding for the problem at hand.

Several significant approximations are made in \Code, notably the
employment of LTE level populations and ionization.  In addition,
bound-bound line transitions are treated using the expansion opacity
formalism (Equation~\ref{Eq:Exp_Opacity}) and an approximate two-level
atom approach to wavelength redistribution.  The ratio of absorption
to scattering in lines is calibrated from detailed atomic models and
is always close to one.  Special care, however, is taken for the
calcium lines, which are assumed to be pure scattering for the reasons
discussed in \S\ref{Sec:Calcium}.  Note that \Code\ allows for a
direct Monte Carlo treatment of line fluorescence, but due to
computational constraints this functionality is not exploited here.
See \cite{Kasen_MC} for further code description and verification.

In the tenuous atmospheres of SNe~Ia, the microscopic conditions
assuring the establishment of LTE level populations and ionization
(namely, the dominance of collisonal rates) are in fact not met. A
direct treatment of the non-LTE physics is thus highly desirable.
Unfortunately, a solution of the non-LTE rate equations including the
$\ga 5$~million potentially important line transitions challenges the
computational capacity of even the most advanced time-independent
spectrum codes.  In a fully time-dependent light curve calculation,
the adoption of LTE at some level appears inevitable.  Fortunately, a
wide range of theoretical calculations confirm the adequacy of LTE in
reproducing the essential spectral and photometric properties of SN~Ia
models in both the optical and NIR
\citep[e.g.,][]{Hoeflich_DD,Baron_NLTE, Pinto-Eastman_II, Wheeler_IR}.
In particular, the crucial iron-group species responsible for
continuum formation have extremely complex atomic structures and
numerous closely spaced atomic levels, and may therefore roughly
approximate their equilibrium distributions.  However, in other atomic
species responsible for individual strong line features non-LTE effects
may be significant (especially calcium, see \S\ref{Sec:Calcium}).  In
addition, because the ionization state of the ejecta is of particular
importance to the NIR lightcurves, LTE becomes increasingly suspect at
later times ($\texp \ga 70$~days), when the non-thermal ionization by
radioactive gamma-rays becomes significant (\S\ref{Sec:Fiducial}).

An additional, perhaps more serious issue for the transfer
calculations is the (in)adequacy of the available atomic data.  Because
NIR emission occurs in millions of relatively weak atomic lines, using
a complete and accurate atomic linelist is critical.  We have found
the commonly applied CD~23 linelist of \cite{Kurucz_1993} (which
contains nearly 500,000 lines) to be insufficient for calculation of
the IJHK-band lightcurves. Upgrading to the Kurucz CD~1 linelist (with
$\sim 42$ million lines), affords significant improvement, however
even this list is likely incomplete and/or inaccurate to some degree.
The inadequacy of the line data at long wavelengths likely plagues our
synthetic NIR lightcurves, as we discuss further in \S\ref{Sec:Fiducial}.

\section{Model NIR Lightcurve of a Normal SNe~Ia}
\label{Sec:Fiducial}

\begin{figure*}
\begin{center}
\includegraphics[width=18.0cm,clip=true]{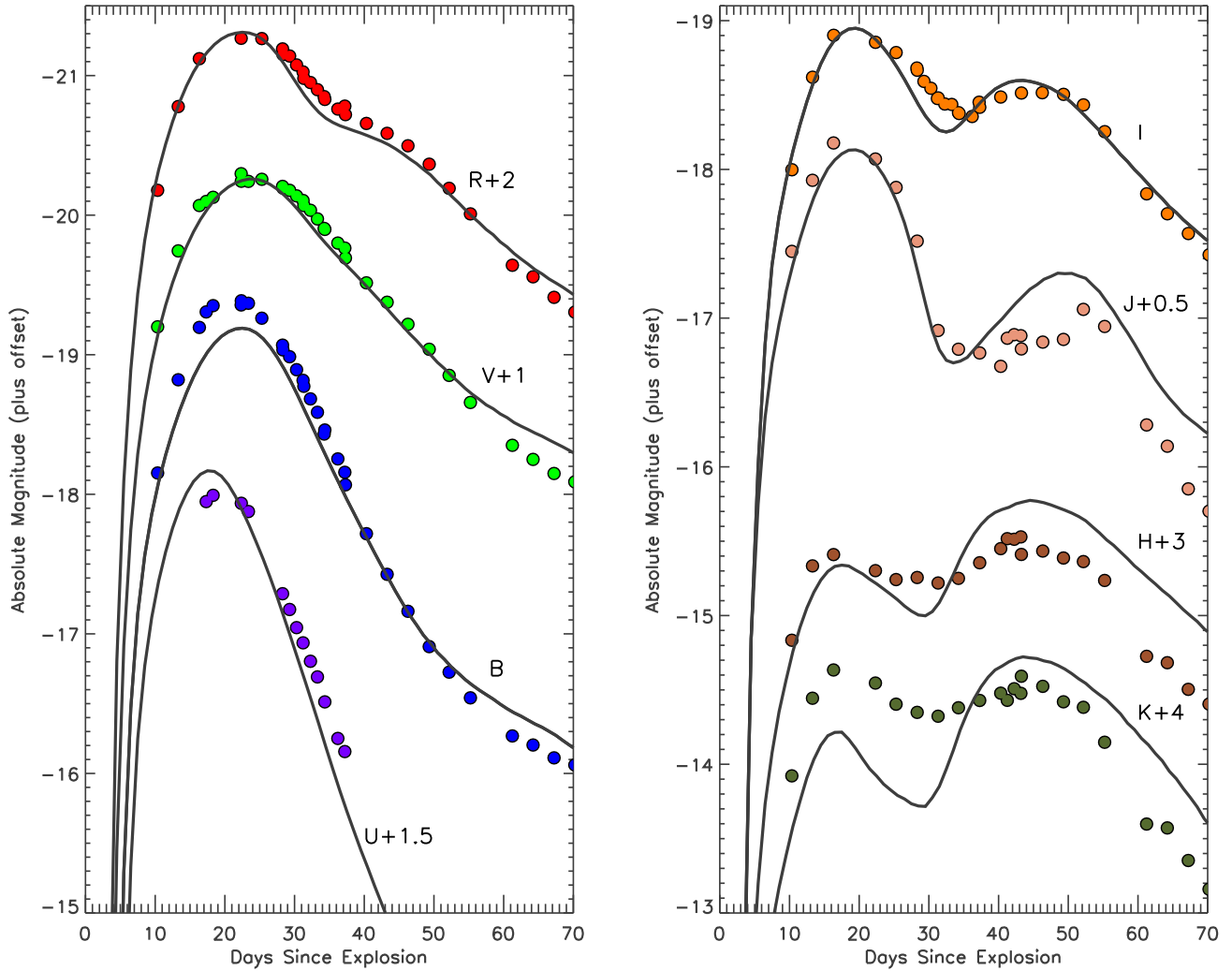}
\caption{Broadband lightcurves of the fiducial model (solid lines)
compared to observations of SN~2001el (filled circles). 
\label{Fig:IRLC}}
\end{center}
\end{figure*}

\begin{figure}
\begin{center}
\includegraphics[width=8.5cm,clip=true]{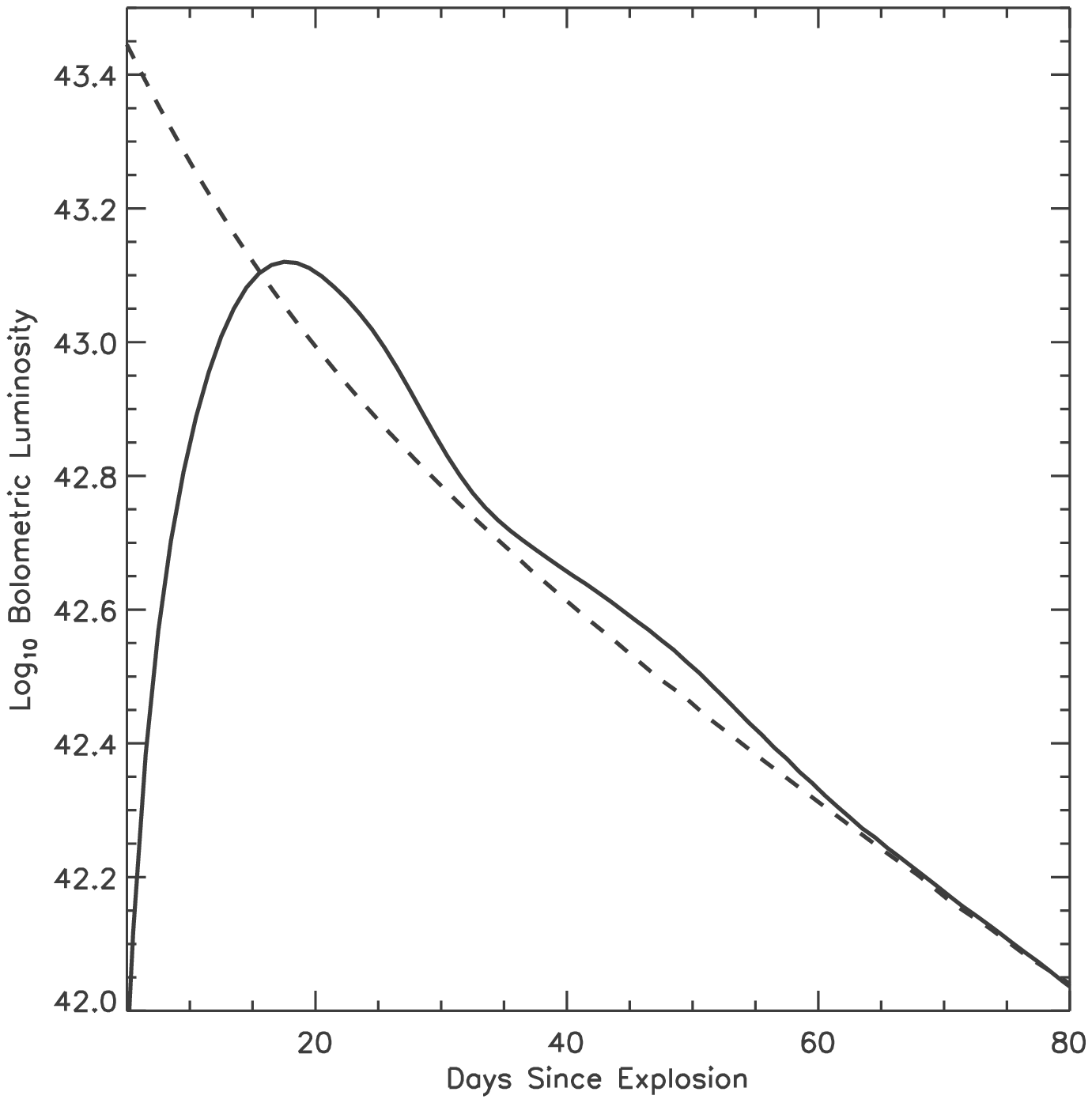}
\caption{Bolometric lightcurve (solid line) and radioactive energy
deposition rate (dashed line) for the fiducial model.  An inflection
in the bolometric lightcurve (corresponding to the NIR secondary
maximum) is seen beginning at day 35 after explosion.
\label{Fig:Bol}}
\end{center}
\end{figure}

We have calculated the full spectral evolution and broadband
lightcurves for our fiducial SN~Ia model containing 0.6 \msun\ of
\Nifs.  Figure~\ref{Fig:IRLC} compares the model optical and NIR
lightcurves to observations of SN~2001el, a photometrically normal
SNe~Ia with extensive coverage obtained by \cite{Kris_01el}.  The
observed lightcurves have been corrected for dust extinction using the
estimates $A_v = 0.57$ \citep{Kris_01el} and $R_v = 2.88$
\citep{Wang_01el}.  We select a distance modulus to SN~2001el of $\mu
= 31.4$ in order to align the peak of the observed $V$-band lightcurve
with that of the model.  This improves the visual comparison of the
lightcurve shapes.  In fact, \cite{Kris_01el} estimate a slightly
smaller distance of $31.29 \pm 0.08$ using the width-luminosity
relation of \cite{Phillips_1999}.  This suggests that our synthetic
lightcurves are perhaps $\sim 0.1$~mag brighter than the observed, and
that a model with slightly lower \Mni\ (e.g., 0.55~\msun) may be more
appropriate for SN~2001el.

The characteristic double-peaked morphology is clearly apparent in all
our model NIR lightcurves, with a distinct and prominent secondary
maximum occurring in the I,J,H, and K-bands.  A corresponding
``shoulder'' is also seen in the R band, and in the bolometric
lightcurve as well (Figure~\ref{Fig:Bol}).  On the whole, the model
V,R,I and J-band lightcurves do a favorable job of fitting those of
SN~2001el, at least for epochs $\texp < 60$~days.  The models get
progressively worse at later times, but this is also when the transfer
calculations are expected to be less reliable.  In the model I-band
lightcurve, the decline from the first maximum is somewhat too rapid
and the secondary maximum occurs slightly earlier than in the
observations.  We note that the prominence and timing of the secondary
maximum is quite sensitive to the model parameters (see
\S\ref{Sec:Depend}) and also shows wide variation in the observations
\citep{Nobili_Iband}.  A thorough exploration of the model parameter
space would likely yield an improved fit to SN~2001el, perhaps in the
process illuminating its particular physical characteristics. In
addition, polarization observations of SN~2001el indicate that the
ejecta was not perfectly spherically symmetric
\citep{Wang_01el,Kasen_01el}, therefore aspherical ejecta structures
may need to be considered to exactly account for all of the lightcurve
properties.

The model does a poorer job in the H and K bands.  In the
observations, the H- and K-band lightcurves are relatively flat for
roughly twenty days after B-band maximum (\texp = 18~days), whereas
the models, in contrast, show a strong rise to the secondary maximum
during these phases.  This is because the model underestimates the
luminosity of the initial maximum relative to the secondary one.
Attempts to improve the H and K fits by varying the model ejecta
structure proved unsuccessful.

\begin{figure}
\begin{center}
\includegraphics[width=8.5cm,clip=true]{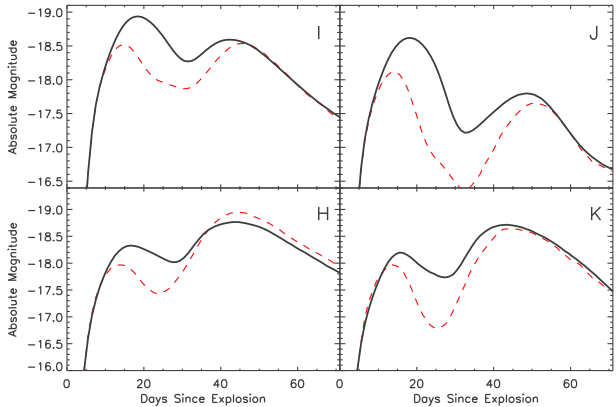}
\caption{Comparison of the IJHK lightcurves of the fiducial model
calculated using two different atomic linelists: the 500,000 line
Kurucz CD 23 (dashed red lines) and the 42 million line Kurucz CD 1
(solid black lines).  Due to the lesser number of lines, the first
maximum of the CD23 lightcurves is depressed relative to CD1 in all
bands.
\label{Fig:Linelist}}
\end{center}
\end{figure}

It is possible that the relatively poor H and K band lightcurve fits
relate to an inadequacy of the LTE approximation, or signal a missing
ingredient in our model ejecta structures.  However, the primary
reason is likely the inadequacy of the atomic linelist.  We
demonstrate the importance of the atomic data by recomputing the
lightcurves using the restricted linelist Kurucz CD23 (containing only
500,000 lines) and comparing to those of our standard linelist (Kurucz
CD1, with 42 million lines).  The differences in the IJHK lightcurves
are considerable (Figure~\ref{Fig:Linelist}), especially at early
times.  This is not surprising, given the large differences noticed in
the opacity calculations of Figure~\ref{Fig:Opacity}

On the whole, the CD23 IJHK lightcurves do a dramatically poorer job
of fitting the observations.  In particular, the initial maximum is
much to weak in all bands.  The problem can be traced directly to the
relative lack of higher ionization, longer wavelength iron group lines
in the CD23 list.  The use of an inadequate linelist may be one reason
why previous theoretical models have sometimes had difficulty fitting
the I-band lightcurves of SNe~Ia.  Although CD1 offers significant
improvement over CD23, it is natural to suspect that were an even more
extensive linelist available, our model $H$ and $K$ band lightcurves
might show significant improvement.

\begin{figure}
\begin{center}
\includegraphics[width=8.5cm,clip=true]{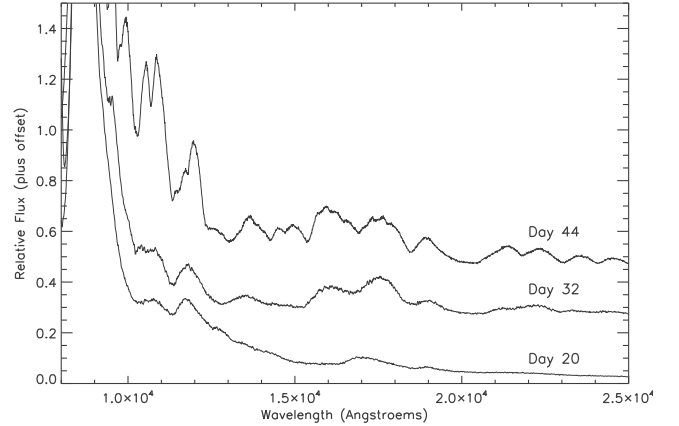}
\caption{Near-infrared spectra of the fiducial SN~Ia model at three
different epochs. Labels indicate time since explosion.
\label{Fig:IR_Spectra}}
\end{center}
\end{figure}

Although not the focus of this paper, we also present NIR spectra of
the fiducial model at a few different epochs in
Figure~\ref{Fig:IR_Spectra}. Near maximum light, the NIR spectrum is
quite featureless, with the occurrence of only a few weak line features
from intermediate mass elements.  At later times, prominent emission
features from blended iron-group lines arise near 1.6 and 1.8~$\mu$.
The spectra in Figure~\ref{Fig:IR_Spectra} show the same general
features identified in the model calculations of \cite{Wheeler_IR}.
The reader can look there for a detailed spectroscopic analysis and
comparison with observations.

\section{Explanation of the Secondary Maximum}
\label{Sec:Explain}

\begin{figure*}[ht]
\begin{center}
\includegraphics[width=18.5cm,clip=true]{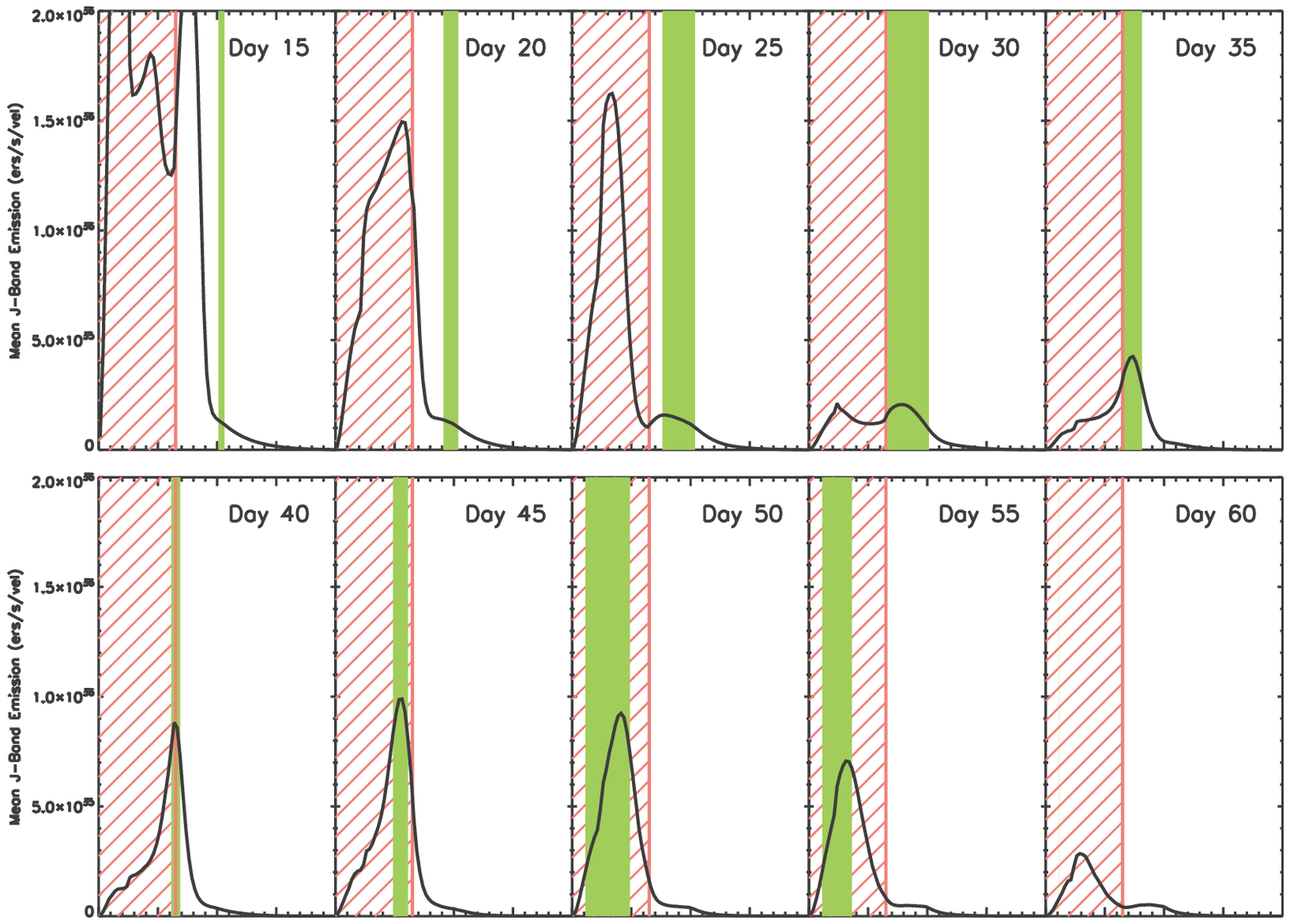}
\caption{Comic strip explaining the origin of the secondary maximum in
the NIR lightcurves of SNe~Ia.  The figure shows the time evolution of
the radial distribution of the instantaneous mean J-band emission
(solid lines).  For each panel, the x-axis is the ejecta velocity
coordinate ranging from zero to 20,000~\kms. The red hatched region
marks the iron core, which extends to $v = 7000$~\kms.  The green
shaded region marks the region near the \rectwoone\ ionization front.
A large spike in J-band emission occurs when that front reaches the
iron rich layers of ejecta, leading to the NIR secondary maximum.
\label{Fig:Emis_Comic}}
\end{center}
\end{figure*}

In our models, we trace the origin of the secondary maximum directly
to the ionization evolution of iron group elements in the ejecta.  In
particular, as discussed in \S\ref{Sec:Emission}, the NIR emissivity
increases sharply at temperature $\Tto \approx 7000$~K in iron/cobalt
gas, marking the transition between the doubly and singly ionized
states (i.e., the \rectwoone\ ionization edge; see
Figure~\ref{Fig:Fe_Emis}).  For reasons already discussed, the
iron-rich gas becomes exceptionally \emph{phosphorescent} at this
temperature, and is very effective in redistributing the pervading
UV/blue radiation to longer wavelengths.

Because the temperature in the ejecta decreases with radius, the
region near the \rectwoone\ ionization front comprises a relative thin
shell of material.  During the hot, early epochs, this region lies in
the far outer layers of ejecta, but as the SN cools overall the
ionization front recedes deeper (in a Lagrangian sense) into the
ejecta.  The resulting inward propagating ``recombination wave''
resembles that often described in models of hydrogen-rich Type~IIP
supernovae.  However, in contrast to the Type~IIP's, the flux mean
opacity in SNe~Ia does not change substantially over the ionization
front and the temperature gradient remains gradual.  It is the onset
of this recombination wave into the iron-rich core of ejecta that
leads to a sudden increase in NIR emission and the re-brightening of
the SN in a secondary maximum.

To delineate this process in detail, we examine the evolution of the
NIR emission over time.  Figure~\ref{Fig:Emis_Comic} is a ``comic
strip'' of the spatial distribution of NIR emissivity (in this case,
J-band) for the fiducial SN~Ia model discussed in the last section.
Each panel shows the radial contribution to the J-band emission at a
given time, specifically $d\psi(r) = \eta_J(r) r^2~dr$ where $r$ is
the radius and $\eta_J$ is the broadband emissivity calculated using
Equation~\ref{Eq:Exp_Emission}.  This figure is meant mainly as a
heuristic -- the observed luminosity in the optically thick phases
depends not only on the instantaneous emission but also the transfer
of radiation through the ejecta.

Given the self-similar nature of the flow, velocity is used as the
radial coordinate in Figure~\ref{Fig:Emis_Comic}.  The red hatched
region identifies the iron-rich core, which extends to roughly
7000~\kms\ in the model.  The green shaded region marks the region
near the \rectwoone\ ionization front (specifically, ionization
fraction between 1.5 and 1.9).  The width of this region, which
depends upon the temperature gradient, is typically a few thousand
\kms.  With time, the ejecta cool and the green shaded region can be
seen moving inward, marking the propagation of the \rectwoone\
recombination wave.

We offer a step-by-step discussion of Figure~\ref{Fig:Emis_Comic}.
During the early epochs (day~15 after explosion), most of the J-band
emission comes from throughout the hot iron core, which has ionization
fraction between 3 and 4.  Note that the emission is always greatest
near the ionization fronts, which are noticed as the emission peaks in
the figure. At these early times, the ejecta is optically thick at NIR
wavelengths to both electron scattering and bound-bound expansion
opacity.  The rise time to the first J-band maximum at \texp = 19~days
is thus related to the diffusion time of photons \citep{Arnett_typeI}.

Following the first J-band maximum, the hot iron core is in the
process of recombining from triply to doubly ionized.  At these times
(days 20 and 25), the NIR emission is dominated by the rather thin
shell of material near the \recthreetwo\ ionization front.  Meanwhile,
the \rectwoone\ ionization front forms well outside the iron core and
contributes little NIR emission.

By day 30 after explosion, the iron core has nearly completely
recombined to doubly ionized and, with the passing of the
\recthreetwo\ ionization front, the NIR emissivity has decreased
significantly.  This is the local minimum in the J-band lightcurve.
The lightcurve would continue to decrease monotonically if it weren't
for the impending onset of the \rectwoone\ recombination wave.
Already this wave has reached the layers just above the iron core,
where a small emission peak is visible.

At day 35 the recombination wave reaches the edge of the iron core,
and the NIR emission begins to increase dramatically.  This is exactly
the time at which the model J-band lightcurve begins to
re-brighten. The NIR emission is now dominated by the ``fluorescent
shell'' at the \rectwoone\ ionization front.  The ejecta are largely
transparent in the NIR and radiation emitted from the shell escapes
almost immediately.  Over the next two weeks, the fluorescent shell
recedes further into the iron core, until, at day 50, it is completely
contained in the iron rich region.  This marks the peak of the J-band
secondary maximum.

At later times, as the fluorescent shell recedes yet deeper in the
ejecta, its volume decreases and so does the J-band brightness.  By
day 60, the \rectwoone\ recombination wave has completely passed, and
the iron core is nearly entirely singly ionized.  The NIR emission is
low and continues to decline.

Why is a secondary maximum seen strongly in the NIR, but not in the
optical lightcurves?  During the epochs of the secondary maximum, the
ejecta remain optically thick at bluer wavelengths ($\lambda \la
5000$~\AA).  The onset of the \rectwoone\ recombination of iron group
elements further increases the opacity in the optical due to the
development of broad blends of Fe~II and Co~II lines in B and
V-bands. Thus optical radiation at the \rectwoone\ ionization front
can not immediately escape the ejecta, but continues to diffuse out
gradually or to fluoresce to longer wavelengths.  Meanwhile, the ejecta
are transparent in the NIR, and photons fluorescing to these
wavelengths escape straightaway.  The net effect of the recombination
wave is therefore to redistribute flux from the optical to longer
wavelengths.

In Figure~\ref{Fig:Bol}, we see that at the onset of the secondary
maximum (day 35 after explosion) a slight bump also occurs in the
model bolometric lightcurve.  This inflection has been noted in
observed SNe~Ia lightcurves \citep{Contardo_bol} and has been explained
by \cite{Pinto-Eastman_II} as the release of pre-existing, trapped
radiative energy.  The ejecta are optically thick (at bluer
wavelengths) for the first couple months after explosion, and the
average diffusion time of photons is a significant fraction of the
expansion time. A store of trapped radiation continually accumulates
in the supernova ejecta.  It is the gradual leaking of this energy
reservoir that causes the bolometric luminosity to exceed the
instantaneous radioactive energy deposition rate from day 18 until
full transparency is reached near day 70.  At day 35, the onset of the
\rectwoone\ ionization front in the iron core suddenly enhances photon
escape via fluorescence into the NIR.  This opens up a new means of
release of the trapped SN radiation, causing a slight rise in the
bolometric lightcurve.

As an interesting aside, notice that the NIR emissivity also peaks
when the gas recombines from singly ionized to neutral
(Figure~\ref{Fig:Fe_Emis}).  In principle, the propagation of another
recombination wave (\reconezero), could lead to a \emph{third} bump in
the NIR lightcurves.  In the current models, a third maximum is in
fact seen at $\texp \approx 100$~days ($\sim 80$ days after B-band
maximum), most prominently in the $J$ band.  However, the inadequacies
of the LTE approximation at late epochs make this behavior suspect.
In particular, the non-thermal ionization from radioactive gamma-rays
becomes significant at low temperatures \citep{Swartz_1991} and will
keep the ejecta from ever going completely neutral.  This may minimize
or delay the putative third maximum.  In addition, emission by
forbidden transitions (not included here) may also become significant
at these epochs. Meanwhile, most published SNe~Ia observations lack
NIR data at times $> 60$ days after maximum.

\section{Dependence on Physical Parameters}
\label{Sec:Depend}

The above explanation of the NIR secondary maximum provides immediate
intuition into the dependence of the NIR lightcurves on the physical
properties of the ejecta.  This allows for simple explanations of the
observed trends concerning the timing and prominence of the secondary
maximum.

First, the rise to a secondary maximum occurs when the \rectwoone\
ionization front first reaches the ejecta layers rich in iron group
elements.  Thus the timing of the secondary maximum will depend upon
both the size of the iron core (or more precisely, its velocity
extent), and the overall temperature scale of the ejecta.  All other
things being equal, SNe with larger iron cores can be expected to have
earlier secondary maxima.  Hotter SNe can be expected to have later
ones.

Second, the luminosity of the second maximum depends fundamentally
upon the size and luminosity of the fluorescent shell formed when the
\rectwoone\ recombination wave has receded into the bulk of the iron
core.  All other things being equal, SNe~Ia with larger iron cores can
be expected to have a brighter secondary maximum.

Below we discuss these behaviors in the context of SNe~Ia ejecta
properties.  In particular, we consider the effects of the degree of
mixing, the \Nifs\ mass, the mass of electron capture elements, the
progenitor metallicity, and the mass of intermediate elements produced,
demonstrating each of these with the models.

\subsection{Mixing}
\label{Sec:Mixing}

\begin{figure}
\begin{center}
\includegraphics[width=8.5cm,clip=true]{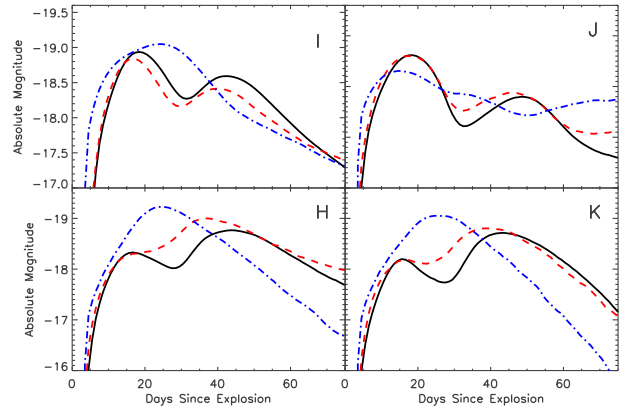}
\caption{Effect of the outward mixing of \Nifs\ on the NIR
lightcurves.  The figure compares the well-stratified fiducial model
(mixing parameter \ani = 0.2, solid black lines) to a model with
enhanced \Nifs\ mixing (\ani = 0.6; red dashed lines) and a model with
a completely homogenized composition structure (blue dot-dashed
lines).
\label{Fig:mixing}}
\end{center}
\end{figure}

The fiducial SN~Ia model studied in the last section was compositionally
stratified, with iron group elements (primarily \Nifs) concentrated in
the central core. The mixing of \Nifs\ outward into the region of IME
effectively increases the velocity extent of the iron core.  Naturally,
we expect this to hasten the occurrence of the NIR secondary maximum.

We demonstrate this effect by comparing the NIR lightcurves of our
fiducial model (mixing parameter $\ani = 0.2$) to one in which \Nifs\
is more thoroughly mixed outward ($\ani = 0.5$).  The result of the
mixing (Figure~\ref{Fig:mixing}) is to advance the NIR secondary
maximum by about 5 days and to decrease its contrast with the first
maximum.  In the I and J-band lightcurves, the first and second maxima
are blended, but still distinguishable.  In the H and K band
lightcurves the double-peaked structure is nearly lost entirely.

In Figure~\ref{Fig:mixing} we show also the lightcurves of a fully
mixed model (i.e., the fiducial model with a completely homogenized
compositional structure).  In this case, the first and second maxima
are indistinguishable in every band (except for perhaps a small
inflection at day 35 in J).  We conclude that the double-peaked
structure observed in NIR lightcurves is a direct indication of the
abundance stratification in SNe~Ia, in particular the concentration of
iron-peak elements in the central regions.  Thus NIR observations
should be useful in constraining the exact degree of mixing in SNe~Ia.

\subsection{Mass of \Nifs}
\label{Sec:Nickel}

\begin{figure}
\begin{center}
\includegraphics[width=8.5cm,clip=true]{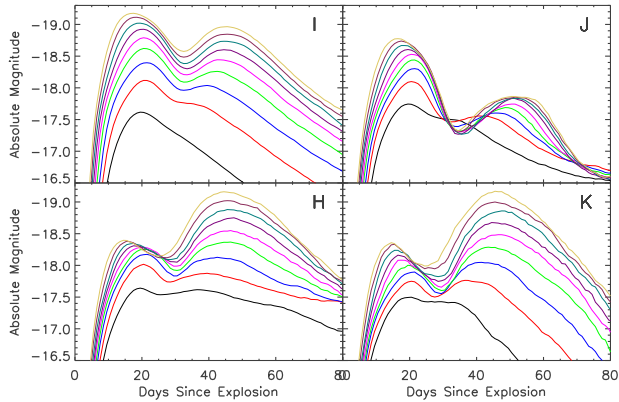}
\caption{Effect of the mass of \Nifs\ (\Mni) on the NIR lightcurves.
The model lightcurves demonstrate variations in \Mni\ from 0.1 to
0.9~\msun\ in 0.1~\msun\ increments.  The less luminous models are
those with lower \Mni.
\label{Fig:Nickel}}
\end{center}
\end{figure}

\begin{figure}[t]
\begin{center}
\includegraphics[width=9cm,clip=true]{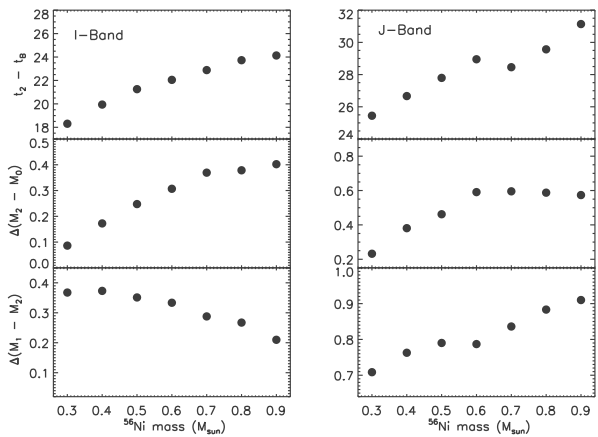}
\caption{Prominence and timing of the secondary maximum versus \Mni\
for the models of Figure~\ref{Fig:Nickel} in the I-band (left) and
J-band (right).  {\it Top panels:} Time (in days) of secondary maximum
minus time of first maximum.  {\it Middle panels:} Number of
magnitudes that the secondary maximum is brighter than the local
minimum that precedes it. {\it Bottom panels:} Number of magnitudes
the first maximum is brighter than the secondary maximum.
\label{Fig:Bump}}
\end{center}
\end{figure}

The mass of \Nifs\ (\Mni) powers the luminosity of SNe~Ia and is the
primary determinate of the peak brightness of the event.  \Mni\ also
affects, in part, the B-band lightcurve decline rate (or stretch) and
is commonly identified as the fundamental parameter controlling the
well-known width-luminosity relation.

In observations, the properties of the I-band secondary maximum are
found to correlate with stretch (and hence peak luminosity), being
more prominent and occurring later in the broader/brighter SNe~Ia.
The origin of these empirical relations can be addressed within the
theoretical paradigm we have described.  To demonstrate, we construct
a series of models (based upon the fiducial model) in which we vary
\Mni\ between 0.l to 0.9~\msun\ in increments of 0.1~\msun.  The sum
of \Mni\ and the mass of IME is held fixed at 1.25~\msun, such that
all models have roughly equal kinetic energy. All other model
parameters have been held fixed.

As in observations, the model secondary maxima are more pronounced in
the higher \Mni\ models, disappearing altogether for the lowest \Mni\
models (Figure~\ref{Fig:Nickel}).  This is due primarily to the larger
size of the iron core in higher \Mni\ models, leading to a larger and
more luminous fluorescent shell at the \rectwoone\ ionization front.
The observed behavior may therefore be taken as further evidence that
more slowly declining SNe~Ia have a larger production of iron group
elements.

\Mni\ produces two opposing effects on the timing of the secondary
maximum.  A larger \Mni\ leads to higher overall temperatures and thus
a later development of the \rectwoone\ recombination wave.  This tends
to delay the secondary maximum.  However, a larger \Mni\ also implies
a larger iron core, which would tend hasten the secondary maximum.
From Figure~\ref{Fig:Nickel}, we see that the first of these effects
dominates, at least for \Mni\ in the range $0.1-0.8$~\msun.  Thus our
models recover the observed timing trend, which should be understood
fundamentally as the physical correlation between the supernova's
luminosity and its ionization state.

For the models with low \Mni, the secondary maximum occurs at early
times and begins merging with the initial maximum.  In our most
subluminous example (\Mni = 0.1~\msun) the ejecta are so cool that the
\rectwoone\ recombination wave reaches the iron core at $\texp \approx
20$~days.  For these objects, the secondary maximum is absent, or
rather it is coincident with the first.  

For reference, Figure~\ref{Fig:Bump} quantifies the timing and
prominence relations in the I and J bands. We note that the observed
trends depend to some extent on how the quantities are empirically
defined.  In the J-band, for example, the peak magnitude of the
secondary maximum increases with \Mni\ relative to the magnitude at the
local minimum, but \emph{decreases} relative to the first maximum.
This reflects complexities glossed over in our discussion (such as the
parameters affecting the initial maximum) and demonstrates the need to
compare observations and models in a uniform way.  Also keep in mind that
additional parameters (described below) can also have profound effects
on the secondary maximum, and may play a role in these relations as
well.

\subsection{Electron Capture Elements and Progenitor Metallicity}
\label{Sec:Iron}

\begin{figure}
\begin{center}
\includegraphics[width=8.5cm,clip=true]{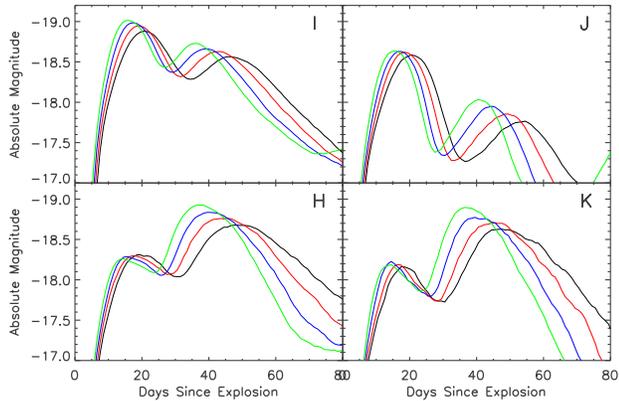}
\caption{Effect of electron capture elements on the NIR lightcurves.
The model lightcurves demonstrate variations in the mass of stable iron
group elements (\Mfe) produced by electron capture at the ejecta
center.  The models are \Mfe = 0.0~\msun\ (black line), \Mfe =
0.1~\msun\ (red line), \Mfe = 0.2~\msun\ (blue line) and \Mfe =
0.3~\msun\ (green line).
\label{Fig:Stretch_Fe}}
\end{center}
\end{figure}

\begin{figure}
\begin{center}
\includegraphics[width=8.5cm,clip=true]{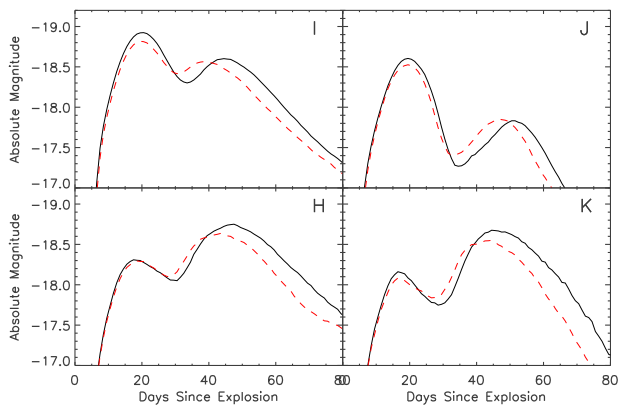}
\caption{Effect of progenitor metallicity on the NIR lightcurves.  The
black line is the fiducial model with zero stable iron production in
the \Nifs\ zone (corresponding to near zero metallicity).  The dashed
red line is the model with 25\% stable iron production throughout the
\Nifs\ zone (corresponding to a metallicity near 3 times solar).
\label{Fig:Metallicity}}
\end{center}
\end{figure}

In addition to \Nifs, SNe~Ia may synthesize substantial amounts of
stable iron group elements, in particular \Feff\ and $^{58}$Ni.
Although this material does not contribute to the overall luminosity
of the SN, it does modify the radial extent of the iron core.  Thus we
anticipate it having important effects on the timing and prominence of
the secondary maximum.

Stable iron group material can be produced by two distinct processes
in SN~Ia explosions: (1) In standard 1-D explosion models, nuclear
burning at the highest densities is subject to electron capture,
leading to the synthesis of 0.5-2.0~\msun\ of neutronized stable iron
group species at the very center of the ejecta
\citep{Nomoto_w7,Thielemann_w7}. There is some observational evidence
suggesting such a pure stable iron center does exist in SNe~Ia
\citep{Hoeflich_ecap}.  (2) While burning to nuclear statistical
equilibrium produces primarily \Nifs, some small fraction of stable
iron group may also be produced, depending upon the metallicity of the
progenitor white dwarf \citep{Thielemann_w7, Timmes_metallicity}.  For
solar metallicity, this fraction would be $\sim 5\%$, while for
metallicity three times solar the fraction may be as high as $\sim
25\%$.  The stable iron group material arising from the progenitor
metallicity would be evenly distributed in low abundance throughout
the \Nifs\ zone.

We demonstrate each of these effects in turn. First, we vary the mass
of stable iron group species (\Mfe) produced by electron capture at
the ejecta center.  We explore values of \Mfe\ from 0 to 3.0~\msun\
while keeping fixed $\Mni = 0.6$~\msun\ and the total burned mass
$\Mfe + \Mni + \Mime = 1.25$~\msun.  Figure~\ref{Fig:Stretch_Fe} shows
the resulting NIR lightcurves.  Because the mass of \Nifs\ is held
fixed, the temperature and ionization evolution is very similar in all
these models.  However, in models with larger values of \Mfe\ (and
hence larger iron cores) the \rectwoone\ recombination wave encounters
the layers of iron rich material at relatively earlier times.  The
secondary maximum thus occurs as many as ten days earlier in models
with larger \Mfe.  In addition, the peak magnitude of the secondary
maximum increases with \Mfe, again due to the larger iron core size.

\cite{Pinto-Eastman_II} have suggested that the stable iron group
material at the ejecta center plays a fundamental role in producing
the NIR secondary maximum.  We note that this is not strictly the
case, as a clear secondary maximum occurs even when no stable iron is
included in the model.

One finds a similar dependence on the stable iron production due to
progenitor metallicity.  In Figure~\ref{Fig:Metallicity} we take the
fiducial model and vary the abundance of stable \Feff\ throughout the
\Nifs\ zone from zero to 25\%, corresponding to a metallicity variation
of roughly zero to 3 times solar. The total mass in the \Nifs\ zone is
held fixed at 0.6~\msun\ -- i.e., in this construction stable iron is
produced at the expense of \Nifs.  Thus the higher metallicity model
has lower bolometric luminosity, lower temperatures, and an earlier
onset of the \rectwoone\ recombination wave.  The secondary maximum
thus occurs about seven days earlier in the higher metallicity model.

Because the size of the iron core remains fixed in the models of
Figure~\ref{Fig:Metallicity}, one expects the absolute luminosity of
the secondary maximum to remains nearly constant.  This is indeed the
case in the I and J band lightcurves. Interestingly, however, the H
and K band secondary maximum are slightly brighter in the low
metallicity models.  This is due to the significantly larger emissivity
of cobalt compared to iron at these wavelengths (see
Figure~\ref{Fig:Comp_Emis}).  Thus the relative strength of the
secondary maxima in different NIR bands may turn out to be a useful
measure of the ratio of cobalt to iron in the bulk ejecta, and hence
the progenitor metallicity.

Significantly, the effect of stable iron group elements on the NIR
lightcurves is distinct from that of \Nifs.  The predicted correlation
(i.e., earlier secondary maxima are as bright or brighter than later
ones) conflicts with the primary observed trend.  Variations in iron
group elements is thus one potential source of deviation from the
standard NIR lightcurve behavior.  This emphasizes the value of NIR
observations in providing constraints of the progenitor metallicity and
the early explosion processes.

\subsection{Calcium Distribution and the IR Triplet Feature}
\label{Sec:Calcium}

\begin{figure}
\begin{center}
\includegraphics[width=8.5cm,clip=true]{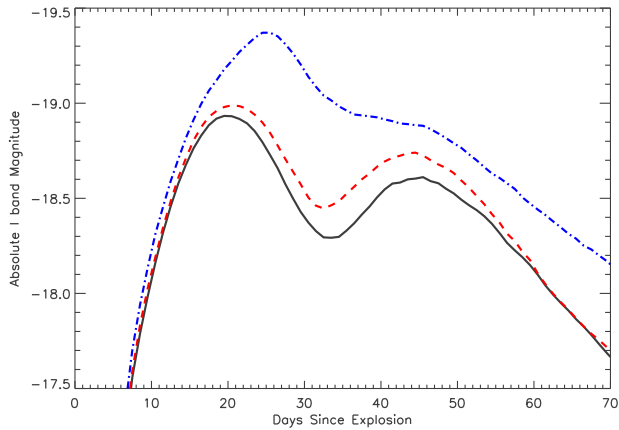}
\caption{Effect of the Ca~II IR triplet feature on the I-band
lightcurve.  For reference, the red dashed line shows the fiducial
model lightcurve when all calcium is removed.  If the calcium lines
are assumed to be pure scattering, the IR triplet feature leads to a
modest decrease in the I-band magnitude (solid black line).  If the
calcium lines are assumed to be pure absorbing, emission in the
triplet feature increases the I-band magnitude significantly (blue
dot-dashed line).
\label{Fig:Calcium}}
\end{center}
\end{figure}

On the whole, the distribution of intermediate mass elements in the
ejecta has little direct effect on the NIR lightcurves.  Emission from
iron group lines dominates the continuum flux, and few individual IME
line features are significant.  The one exception is in the I-band,
where the prominent Ca~II IR triplet feature ($gf$-weighted rest
wavelength 8579~\AA) can potentially affect the photometry
significantly.  The triplet lines are intrinsically very strong, and may
remain optically thick far out in the low density layers of ejecta,
even when the calcium abundance is very low ($\Xca \approx 0.01$).

Unfortunately, the net effect of the Ca~II lines on the I-band
lightcurve depends upon the details of the radiative transfer,
specifically the line source functions.  Two limiting cases can be
considered: (1) If the calcium lines are assumed to be pure
scattering, flux is nearly conserved over the line profile -- i.e.,
the absorption component of the P-Cygni profile nearly cancels the
emission component.  In this case, the line feature should have only a
minor effect on the broadband magnitude. (2) If, on the other hand,
the calcium lines are assumed to be completely absorbing (as in pure
LTE calculations) the net emission in the Ca~II IR triplet feature can
be considerable, because radiation absorbed in blue part of the
spectrum is allowed to re-emerge in the triplet lines.  In the models
of this paper we have always assumed pure scattering calcium lines,
for the reasons described below.

Figure~\ref{Fig:Calcium} quantifies the net effect of calcium on the
I-band lightcurve by comparing the fiducial model to a similar model
in which all calcium has been removed.  Under the pure scattering
assumption, the Ca~II triplet feature causes modest ($\sim 0.1$~mag)
I-band differences. These arise for two reasons.  First, because of
blending with other lines, flux is not strictly conserved over the
triplet line profile.  Second, because the \cite{Bessell_1990} I-band
filter profile cuts off near 8500~\AA, the absorption and emission
component of the triplet feature are not equally sampled.  Thus, under
pure scattering, the calcium feature serves to slightly
\emph{decrease} the I-band magnitude.

If, on the other hand, the calcium lines are assumed to be pure
absorbing, net emission in the triplet features enhances the peak
I-band magnitude by $\sim 0.5$~mag, and delays the time of peak by
nearly seven days.  The secondary maximum is no longer distinct and
prominent, and the model I-band lightcurve does not fit the observed
lightcurve shape or colors.  The result emphasizes that the I-band
secondary maximum does not occur because of emission in the Ca~II
triplet feature, but rather \emph{in spite} of it.

From an atomic physics perspective, the low collisonal rates in the SN
ejecta and the large oscillator strength of the Ca~II triplet lines
suggest that pure scattering is indeed the more plausible
representation of the line source functions.  This is supported by our
lightcurve fits to the SN~2001el observations.  To the extent that
pure scattering does hold, the shape and peak brightness of the
I-band lightcurves of SNe~Ia will be generally insensitive (at the
$\la 0.1$~mag level) to the details of calcium production and
distribution.  This is an important condition for the robust
application of I-band lightcurves to cosmology studies.  One
interesting exception is worth noting: if the calcium distribution
deviates from spherical symmetry, flux scattered in the triple feature
will be redistributed anisotropically along different viewing angles.

The results of this section necessitate the inclusion of a non-LTE
treatment of the calcium level populations within the fully
time-dependent multi-dimensional radiative transfer calculation.  This
solution is in fact readily incorporated into the Monte Carlo
approach, and will be included in the upcoming version of \Code.  The
common adoption of purely absorbing lines in previous LTE radiative
transfer studies may be another reason why those calculations often
found difficulty in fitting the I-band lightcurves of SNe~Ia.

\section{Near-IR SN~Ia As Standard Candles}
\label{Sec:Dispersion}

\begin{figure}
\begin{center}
\includegraphics[width=8.5cm,clip=true]{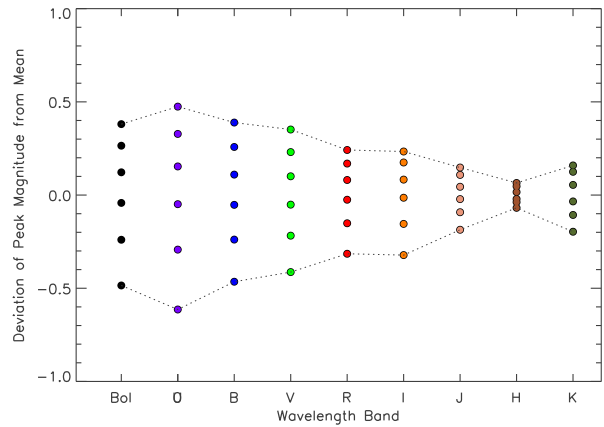}
\caption{Dispersion in peak magnitude (measured at the first
lightcurve maximum) as a function of wavelength band for the models
of Figure~\ref{Fig:Nickel} with \Nifs\ masses between 0.4 and
0.9~\msun.
\label{Fig:Dispersion}}
\end{center}
\end{figure}

The cosmological interest in NIR observations of SNe~Ia stems from the
small observed dispersion ($\sim 0.2$~mag) in the peak JHK-band
magnitudes.  We explore this property here using the models discussed
in \S\ref{Sec:Nickel} (and shown in Figure~\ref{Fig:Nickel}) for which
the mass of \Nifs\ is varied.  We consider only those models with
\Nifs\ mass in the ``normal'' range of 0.4 to 0.9~\msun.

Figure~\ref{Fig:Dispersion} shows the spread in model peak magnitude
(measured at the first maximum) as a function of broadband filter.
One notices a general decline in the dispersion towards the red.  In
the B-band, the spread is roughly $\pm 0.5$~mag, similar to the
observed (uncorrected) dispersion \citep{Hamuy_96b}. This decreases to
$\pm 0.4$~mag in V and $\pm 0.3$~mag in R and I.  In the NIR, the
dispersion is more than a factor two in magnitude smaller than the
B-band, being $\sim 0.2$~mag in J and K and $\sim 0.1$~mag in H.  For
the H and K bands, bear in mind the failure of the models to fix the
observed first maximum, presumed an inadequacy of the atomic data.
Nevertheless, this level of NIR dispersion is very similar to that
inferred from observations \citep{Meikle_IR2000}.

In addition, the model peak NIR magnitudes show an impressive
insensitivity to the other physical parameters explored in this
paper. Significant variations in the progenitor metallicity, electron
capture elements, and IME production all only led to small ($\sim
0.1$~mag) variations in the absolute magnitude of the initial peak
(Figures~\ref{Fig:Stretch_Fe}, \ref{Fig:Metallicity}, and
\ref{Fig:Calcium}).  The mixing of \Nifs\ also caused only slight
magnitude variations unless extreme levels of mixing were considered
(Figure~\ref{Fig:mixing}).  These results further testify to the robustness
of SNe~Ia NIR lightcurves as standard candles.

Our model lightcurves may also suggest further refinements to
cosmology studies in the NIR.  The J-band lightcurves, for example,
show almost \emph{zero} intrinsic dispersion at the local minimum
occurring between the first and secondary maximum (about 15 days after
B-band maximum; see Figure~\ref{Fig:Nickel}).  Even when the
subluminous models are considered the magnitude variations at this
epoch are small.  It would be interesting to check if observed SNe~Ia
exhibit a similar behavior.  As it turns out, \cite{Meikle_IR2000}
chose to study the J-band magnitudes of SNe~Ia measured 14 days after
B-maximum (as derived from fits to the \cite{Elias_1985} templates).
He found that the dispersion in $M_{14}(J)$ was small, although not
necessarily minimal.

The low peak magnitude dispersion in the NIR follows from the
dependence of the model colors on the bolometric luminosity.  The
dimmer SNe also have lower temperatures, and radiate a greater
percentage of the energy at redder wavelengths.  This acts as a
regulating mechanism which maintains a nearly constant peak magnitude
in the NIR bands, regardless of the \Nifs\ mass.  On the other hand,
the dispersion in ultraviolet and $U$-band magnitudes is reciprocally
intensified, exceeding that of the bolometric lightcurves.

\section{Discussion and Conclusion}

We have modeled the far-red and NIR lightcurves of SNe~Ia and given a
detailed explanation of the characteristic secondary maximum.  Our
synthetic model lightcurves were calculated using parameterized 1-D
ejecta configurations and the time-dependent multi-group radiative
transfer code \Code.  The model lightcurves displayed distinct and
conspicuous secondary maxima and provided favorable fits to the NIR
observations of the normal Type~Ia SN~2001el.  By varying the mass of
\Nifs, the models also reproduced the observed trend that brighter
SNe~Ia have later and more prominent secondary maxima.

We trace the origin of the secondary maximum directly to the
ionization evolution of iron group elements in the ejecta.
Specifically, the NIR emissivity of iron/cobalt gas peaks sharply at a
temperature $\Tto \approx 7000$~K, marking the transition between the
singly and doubly ionized states.  The recombination of iron-rich gas
from \rectwoone\ thus leads to enhanced redistribution of radiation
from blue to NIR.  Interestingly, as the supernova cools, the global
ionization evolution takes the form a \rectwoone\ ``recombination
wave'' gradually receding deeper into the ejecta.  The onset and
propagation of this wave through the iron rich layers marks the rise
and fall the secondary maximum.  Because the ejecta are transparent at
longer wavelengths during these epochs, the NIR observations allow us
to watch directly as the recombination wave scans progressively
through deeper and deeper layers of ejecta.

While the models considered in this paper captured the essential
aspects of SN~Ia NIR lightcurves, they also highlighted several
outstanding issues for the radiative transfer calculations.  First,
the use of a complete and accurate atomic linelist proved critical in
modeling the NIR bands.  Further improvement of the currently
available atomic data is likely needed to accurately model the H and
K-band lightcurves.  Second, non-thermal ionization effects from
radioactive gamma-rays become significant for $\texp \ga 70$~days when
LTE predicts neutrality, and thus likely have a dramatic impact on the
NIR lightcurves at these late epochs.  Third, a proper treatment of
Ca~II IR triplet line source function is crucial in synthesizing
accurate I-band model lightcurves, as the assumption of purely
absorbing lines leads to unrealistic results.  Future advances to
the \Code\ code will permit more detailed studies of these and other
important effects.  However, one does not expect the technical
developments to change the general NIR lightcurve trends and
dependencies explained in this paper.

In this paper, we studied the dependence of the NIR lightcurves on a
number of important physical parameters, which highlighted the many
ways in which NIR lightcurves offer valuable diagnostics of the
ejecta properties and powerful constraints on explosion models.

First, the double-peaked morphology of the NIR lightcurves can be
taken as a direct consequence of the abundance stratification in
SNe~Ia, in particular the concentration of iron group elements in the
central regions.  This confirms previous inferences based upon
post-maximum and nebular spectra \citep[e.g.,][]{Branch_w7,
Hoeflich_99by, Kozma_neb}.  Abundance stratification is generic to
certain classes of explosion models, for example the delayed
detonation models \citep{Khokhlov_DD} and the detonation from failed
detonation \citep{Plewa_GCD}.  In contrast, models characterized by
large-scale mixing, such as published 3-dimensional deflagration
models \citep{Gamezo_3D,Reinecke_3D} are likely inconsistent with the
double peaked behavior in the NIR lightcurves.  Further NIR
observations, coupled with the transfer models, should be useful in
constraining the exact degree of \Nifs\ mixing in SNe~Ia, and thus in
testing current and future explosion paradigms.

Second, the luminosity of the secondary maximum provides a measure of
the amount of iron group elements (both stable and radioactive)
synthesized in the explosion.  The observed correlation between B-band
decline rate and the luminosity of the secondary maximum provides
strong evidence that slower declining SNe~Ia have (on average) a
larger production of iron group elements.  This conclusion coincides
with several other inferences to the same \citep{Contardo_bol,
Mazzali_nebular, Stritzinger}.

Third, the NIR secondary maximum is sensitive to the amount of stable
iron group elements produced in the explosion, and hence the
progenitor metallicity.  \cite{Timmes_metallicity} has suggested that
metallicity variations may lead to 25\% variations in \Nifs.  We find
that the NIR signature of this variation is distinct from that of
varying \Nifs\ independently.  Our predicted correlation arising from
metallicity variations is that earlier secondary maxima will be as
bright or brighter than later ones.  This conflicts with the primary
observed trend, and suggests that the metallicity is likely a
sub-dominant cause of SN~Ia luminosity variations.  This signature
further opens the possibility of using an observational sample of NIR
lightcurves to constrain metallicity variations among different
progenitor populations.

Fourth, the timing of the secondary maximum is a direct probe of the
temperature and ionization evolution in the ejecta, and hence a useful
tool in interpreting observations.  As it turns out, the ionization
evolution of crucial importance to the optical lightcurve decline rate
as well. The \rectwoone\ recombination of iron group elements
contributes significantly to blanketing in the B and V-bands due to the
development of blends of strong Fe~II and Co~II lines in the optical
part of the spectrum. In this context, NIR observations may also be
especially useful in diagnosing peculiar objects.  For example, the
paradoxical SN~2002cx was $\sim 2$ mag subluminous, but had a peculiar
spectrum characteristic of ``hot'' iron-rich ejecta \citep{Li_02cx}.
NIR observations would compliment that spectroscopic analyses be
providing independent measures of the temperature scale, ionization
evolution, iron core size and mixing.

Finally, we reiterate the cosmological utility of the NIR lightcurves
of SNe~Ia.  In our lightcurve models, the dispersion in NIR peak
magnitudes is found to be small ($\la 0.2$~mag) even when the physical
properties of the ejecta are varied widely.  At certain epochs (e.g.,
the J-band local minimum occurring near day 15 after B-maximum) the
predicted dispersion is even smaller.  Our models thus further testify
to the robustness of SNe~Ia NIR lightcurves as standard candles.

Perhaps more importantly, the NIR observations provide deep and
readily interpreted diagnostics of the physical conditions in the
SN~Ia ejecta, and therefore offer an opportunity for controlling
potential systematic errors in the cosmology studies.  We have shown
that the physical parameters affecting the NIR secondary maximum are
also those that may modulate the optical peak brightness and decline
rate of SNe~Ia.  Thus the NIR lightcurves are one of the most
promising places to search for empirical secondary parameters related
to deviations from the standard width-luminosity relation.

In this context, the NIR observations strike a nice balance in their
level of information content.  Given the double-peaked morphology, the
NIR lightcurves carry considerably more information than do the
single-peaked optical band observations; at the same time, they remain
simple enough to submit to a much more ready statistical analysis than
do the complex spectral time series.  For all these reasons, a sizable
sample of observed NIR lightcurves would be a remarkably powerful tool
in constraining the possible evolution or progenitor drift of SNe~Ia
arising in differing stellar populations.  This potential should be
exploited further with expanded observational programs and concerted
theoretical efforts.

\acknowledgements The author thanks Kevin Krisciunas and Peter Meikle
for helpful comments and suggestions.


\begin{thebibliography}{54}
\expandafter\ifx\csname natexlab\endcsname\relax\def\natexlab#1{#1}\fi

\bibitem[{{Arnett}(1982)}]{Arnett_typeI}
{Arnett}, W.~D. 1982, ApJ, 253, 785

\bibitem[{{Baron} {et~al.}(1996){Baron}, {Hauschildt}, {Nugent}, \&
  {Branch}}]{Baron_NLTE}
{Baron}, E., {Hauschildt}, P.~H., {Nugent}, P., \& {Branch}, D. 1996, MNRAS,
  283, 297

\bibitem[{{Bessell}(1990)}]{Bessell_1990}
{Bessell}, M.~S. 1990, PASP, 102, 1181

\bibitem[{{Blinnikov} \& {Sorokina}(2004)}]{Blinnikov_2004}
{Blinnikov}, S., \& {Sorokina}, E. 2004, Ap\&SS, 290, 13

\bibitem[{{Bowers} {et~al.}(1997){Bowers}, {Meikle}, {Geballe}, {Walton},
  {Pinto}, {Dhillon}, {Howell}, \& {Harrop-Allin}}]{Bowers_86G}
{Bowers}, E.~J.~C., {Meikle}, W.~P.~S., {Geballe}, T.~R., {Walton}, N.~A.,
  {Pinto}, P.~A., {Dhillon}, V.~S., {Howell}, S.~B., \& {Harrop-Allin}, M.~K.
  1997, MNRAS, 290, 663

\bibitem[{{Branch} {et~al.}(1985){Branch}, {Doggett}, {Nomoto}, \&
  {Thielemann}}]{Branch_w7}
{Branch}, D., {Doggett}, J.~B., {Nomoto}, K., \& {Thielemann}, F.-K. 1985, ApJ,
  294, 619

\bibitem[{{Candia} {et~al.}(2003){Candia}, {Krisciunas}, {Suntzeff},
  {Gonz{\'a}lez}, {Espinoza}, {Leiton}, {Rest}, {Smith}, {Cuadra}, {Tavenner},
  {Logan}, {Snider}, {Thomas}, {West}, {Gonz{\'a}lez}, {Gonz{\'a}lez},
  {Phillips}, {Hastings}, \& {McMillan}}]{Candia_00cx}
{Candia}, P. {et~al.} 2003, PASP, 115, 277

\bibitem[{{Contardo} {et~al.}(2000){Contardo}, {Leibundgut}, \&
  {Vacca}}]{Contardo_bol}
{Contardo}, G., {Leibundgut}, B., \& {Vacca}, W.~D. 2000, A\&A, 359, 876

\bibitem[{{Eastman} \& {Pinto}(1993)}]{Pinto-Eastman_Spectra}
{Eastman}, R.~G., \& {Pinto}, P.~A. 1993, ApJ, 412, 731

\bibitem[{{Elias} {et~al.}(1981){Elias}, {Frogel}, {Hackwell}, \&
  {Persson}}]{Elias_1981}
{Elias}, J.~H., {Frogel}, J.~A., {Hackwell}, J.~A., \& {Persson}, S.~E. 1981,
  ApJ, 251, L13

\bibitem[{{Elias} {et~al.}(1985){Elias}, {Matthews}, {Neugebauer}, \&
  {Persson}}]{Elias_1985}
{Elias}, J.~H., {Matthews}, K., {Neugebauer}, G., \& {Persson}, S.~E. 1985,
  ApJ, 296, 379

\bibitem[{{Gamezo} {et~al.}(2003){Gamezo}, {Khokhlov}, {Oran}, {Chtchelkanova},
  \& {Rosenberg}}]{Gamezo_3D}
{Gamezo}, V.~N., {Khokhlov}, A.~M., {Oran}, E.~S., {Chtchelkanova}, A.~Y., \&
  {Rosenberg}, R.~O. 2003, Science, 299, 77

\bibitem[{{H{\" o}flich} {et~al.}(2002){H{\" o}flich}, {Gerardy}, {Fesen}, \&
  {Sakai}}]{Hoeflich_99by}
{H{\" o}flich}, P., {Gerardy}, C.~L., {Fesen}, R.~A., \& {Sakai}, S. 2002, ApJ,
  568, 791

\bibitem[{{Hamuy} {et~al.}(1996{\natexlab{a}}){Hamuy}, {Phillips}, {Suntzeff},
  {Schommer}, {Maza}, \& {Aviles}}]{Hamuy_96b}
{Hamuy}, M., {Phillips}, M.~M., {Suntzeff}, N.~B., {Schommer}, R.~A., {Maza},
  J., \& {Aviles}, R. 1996{\natexlab{a}}, AJ, 112

\bibitem[{{Hamuy} {et~al.}(1996{\natexlab{b}}){Hamuy}, {Phillips}, {Suntzeff},
  {Schommer}, {Maza}, {Smith}, {Lira}, \& {Aviles}}]{Hamuy_96Templates}
{Hamuy}, M., {Phillips}, M.~M., {Suntzeff}, N.~B., {Schommer}, R.~A., {Maza},
  J., {Smith}, R.~C., {Lira}, P., \& {Aviles}, R. 1996{\natexlab{b}}, AJ, 112,
  2438

\bibitem[{{Hernandez} {et~al.}(2000){Hernandez}, {Meikle}, {Aparicio}, {Benn},
  {Burleigh}, {Chrysostomou}, {Fernandes}, {Geballe}, {Hammersley},
  {Iglesias-Paramo}, {James}, {James}, {Kemp}, {Lister}, {Martinez-Delgado},
  {Oscoz}, {Pollacco}, {Rozas}, {Smartt}, {Sorensen}, {Swaters}, {Telting},
  {Vacca}, {Walton}, \& {Zapatero-Osorio}}]{Hernandez_98bu}
{Hernandez}, M. {et~al.} 2000, MNRAS, 319, 223

\bibitem[{{Hoeflich} \& {Khokhlov}(1996)}]{Hoeflich_Khokhlov_LC}
{Hoeflich}, P., \& {Khokhlov}, A. 1996, ApJ, 457, 500

\bibitem[{{Hoflich}(1995)}]{Hoeflich_94D}
{Hoflich}, P. 1995, ApJ, 443, 89

\bibitem[{{H{\"o}flich} {et~al.}(2004){H{\"o}flich}, {Gerardy}, {Nomoto},
  {Motohara}, {Fesen}, {Maeda}, {Ohkubo}, \& {Tominaga}}]{Hoeflich_ecap}
{H{\"o}flich}, P., {Gerardy}, C.~L., {Nomoto}, K., {Motohara}, K., {Fesen},
  R.~A., {Maeda}, K., {Ohkubo}, T., \& {Tominaga}, N. 2004, ApJ, 617, 1258

\bibitem[{{Hoflich} {et~al.}(1995){Hoflich}, {Khokhlov}, \&
  {Wheeler}}]{Hoeflich_DD}
{Hoflich}, P., {Khokhlov}, A.~M., \& {Wheeler}, J.~C. 1995, ApJ, 444, 831

\bibitem[{{Karp} {et~al.}(1977){Karp}, {Lasher}, {Chan}, \& {Salpeter}}]{Karp}
{Karp}, A.~H., {Lasher}, G., {Chan}, K.~L., \& {Salpeter}, E.~E. 1977, \apj,
  214, 161

\bibitem[{{Kasen} {et~al.}(2003){Kasen}, {Nugent}, {Wang}, {Howell}, {Wheeler},
  {H{\" o}flich}, {Baade}, {Baron}, \& {Hauschildt}}]{Kasen_01el}
{Kasen}, D. {et~al.} 2003, ApJ, 593, 788

\bibitem[{{Kasen} {et~al.}(2006)}]{Kasen_MC}
{Kasen}, D., {et~al.} 2006, ApJ, submitted

\bibitem[{Khokhlov(1991)}]{Khokhlov_DD}
Khokhlov, A. 1991, A\&A, 245, 114

\bibitem[{{Kirshner} {et~al.}(1973){Kirshner}, {Willner}, {Becklin},
  {Neugebauer}, \& {Oke}}]{Kirshner_1973}
{Kirshner}, R.~P., {Willner}, S.~P., {Becklin}, E.~E., {Neugebauer}, G., \&
  {Oke}, J.~B. 1973, ApJ, 180, L97+

\bibitem[{{Kozma} {et~al.}(2005){Kozma}, {Fransson}, {Hillebrandt},
  {Travaglio}, {Sollerman}, {Reinecke}, {R{\"o}pke}, \&
  {Spyromilio}}]{Kozma_neb}
{Kozma}, C., {Fransson}, C., {Hillebrandt}, W., {Travaglio}, C., {Sollerman},
  J., {Reinecke}, M., {R{\"o}pke}, F.~K., \& {Spyromilio}, J. 2005, A\&A, 437,
  983

\bibitem[{{Krisciunas} {et~al.}(2001){Krisciunas}, {Phillips}, {Stubbs},
  {Rest}, {Miknaitis}, {Riess}, {Suntzeff}, {Roth}, {Persson}, \&
  {Freedman}}]{Kris_2001AJ}
{Krisciunas}, K. {et~al.} 2001, AJ, 122, 1616

\bibitem[{{Krisciunas} {et~al.}(2004{\natexlab{a}}){Krisciunas}, {Phillips}, \&
  {Suntzeff}}]{Kris_IRHubble}
{Krisciunas}, K., {Phillips}, M.~M., \& {Suntzeff}, N.~B. 2004{\natexlab{a}},
  ApJ, 602, L81

\bibitem[{{Krisciunas} {et~al.}(2004{\natexlab{b}}){Krisciunas}, {Phillips},
  {Suntzeff}, {Persson}, {Hamuy}, {Antezana}, {Candia}, {Clocchiatti}, {DePoy},
  {Germany}, {Gonzalez}, {Gonzalez}, {Krzeminski}, {Maza}, {Nugent}, {Qiu},
  {Rest}, {Roth}, {Stritzinger}, {Strolger}, {Thompson}, {Williams}, \&
  {Wischnjewsky}}]{Kris_2004AJ2}
{Krisciunas}, K. {et~al.} 2004{\natexlab{b}}, AJ, 127, 1664

\bibitem[{{Krisciunas} {et~al.}(2003){Krisciunas}, {Suntzeff}, {Candia},
  {Arenas}, {Espinoza}, {Gonzalez}, {Gonzalez}, {H{\"o}flich}, {Landolt},
  {Phillips}, \& {Pizarro}}]{Kris_01el}
---. 2003, AJ, 125, 166

\bibitem[{{Krisciunas} {et~al.}(2004{\natexlab{c}}){Krisciunas}, {Suntzeff},
  {Phillips}, {Candia}, {Prieto}, {Antezana}, {Chassagne}, {Chen}, {Dickinson},
  {Eisenhardt}, {Espinoza}, {Garnavich}, {Gonz{\'a}lez}, {Harrison}, {Hamuy},
  {Ivanov}, {Krzemi{\'n}ski}, {Kulesa}, {McCarthy}, {Moro-Mart{\'{\i}}n},
  {Muena}, {Noriega-Crespo}, {Persson}, {Pinto}, {Roth}, {Rubenstein},
  {Stanford}, {Stringfellow}, {Zapata}, {Porter}, \&
  {Wischnjewsky}}]{Kris_2004AJ1}
---. 2004{\natexlab{c}}, AJ, 128, 3034

\bibitem[{Kurucz(1993)}]{Kurucz_1993}
Kurucz, R. 1993, CD-ROM 1, Atomic Data for Opacity Calculations (Cambridge:
  Smithsonian Astrophysical Observatory)

\bibitem[{{Li} {et~al.}(2003){Li}, {Filippenko}, {Chornock}, {Berger},
  {Berlind}, {Calkins}, {Challis}, {Fassnacht}, {Jha}, {Kirshner}, {Matheson},
  {Sargent}, {Simcoe}, {Smith}, \& {Squires}}]{Li_02cx}
{Li}, W. {et~al.} 2003, PASP, 115, 453

\bibitem[{{Marion} {et~al.}(2003){Marion}, {H{\"o}flich}, {Vacca}, \&
  {Wheeler}}]{Marion_IR}
{Marion}, G.~H., {H{\"o}flich}, P., {Vacca}, W.~D., \& {Wheeler}, J.~C. 2003,
  ApJ, 591, 316

\bibitem[{{Mazzali} {et~al.}(1998){Mazzali}, {Cappellaro}, {Danziger},
  {Turatto}, \& {Benetti}}]{Mazzali_nebular}
{Mazzali}, P.~A., {Cappellaro}, E., {Danziger}, I.~J., {Turatto}, M., \&
  {Benetti}, S. 1998, ApJ, 499, L49

\bibitem[{{Mazzali} \& {Lucy}(1993)}]{Mazzali_MC}
{Mazzali}, P.~A., \& {Lucy}, L.~B. 1993, A\&A, 279, 447

\bibitem[{{Meikle}(2000)}]{Meikle_IR2000}
{Meikle}, W.~P.~S. 2000, MNRAS, 314, 782

\bibitem[{Meikle {et~al.}(1996)}]{Meikle_94D}
Meikle, W. P.~S., {et~al.} 1996, MNRAS, 281, 263

\bibitem[{Mihalas(1978)}]{Mihalas_SA}
Mihalas, D. 1978, Stellar Atmospheres (San Francisco: W. H. Freeman)

\bibitem[{{Nobili} {et~al.}(2005){Nobili}, {Amanullah}, {Garavini}, {Goobar},
  {Lidman}, {Stanishev}, {Aldering}, {Antilogus}, {Astier}, {Burns}, {Conley},
  {Deustua}, {Ellis}, {Fabbro}, {Fadeyev}, {Folatelli}, {Gibbons}, {Goldhaber},
  {Groom}, {Hook}, {Howell}, {Kim}, {Knop}, {Nugent}, {Pain}, {Perlmutter},
  {Quimby}, {Raux}, {Regnault}, {Ruiz-Lapuente}, {Sainton}, {Schahmaneche},
  {Smith}, {Spadafora}, {Thomas}, {Wang}, \& {The Supernova Cosmology
  Project}}]{Nobili_Iband}
{Nobili}, S. {et~al.} 2005, \aap, 437, 789

\bibitem[{Nomoto {et~al.}(1984)Nomoto, Thielemann, \& Yokoi}]{Nomoto_w7}
Nomoto, K., Thielemann, F., \& Yokoi, K. 1984, ApJ, 286, 644

\bibitem[{{Persson} {et~al.}(1998){Persson}, {Murphy}, {Krzeminski}, {Roth}, \&
  {Rieke}}]{Persson_IR}
{Persson}, S.~E., {Murphy}, D.~C., {Krzeminski}, W., {Roth}, M., \& {Rieke},
  M.~J. 1998, AJ, 116, 2475

\bibitem[{{Phillips} {et~al.}(1999){Phillips}, {Lira}, {Suntzeff}, {Schommer},
  {Hamuy}, \& {Maza}}]{Phillips_1999}
{Phillips}, M.~M., {Lira}, P., {Suntzeff}, N.~B., {Schommer}, R.~A., {Hamuy},
  M., \& {Maza}, J. 1999, AJ, 118, 1766

\bibitem[{{Pinto} \& {Eastman}(2000)}]{Pinto-Eastman_II}
{Pinto}, P.~A., \& {Eastman}, R.~G. 2000, ApJ, 530, 757

\bibitem[{{Plewa} {et~al.}(2004){Plewa}, {Calder}, \& {Lamb}}]{Plewa_GCD}
{Plewa}, T., {Calder}, A.~C., \& {Lamb}, D.~Q. 2004, ApJ, 612, L37

\bibitem[{{Reinecke} {et~al.}(2002){Reinecke}, {Hillebrandt}, \&
  {Niemeyer}}]{Reinecke_3D}
{Reinecke}, M., {Hillebrandt}, W., \& {Niemeyer}, J.~C. 2002, A\&A, 391, 1167

\bibitem[{{Rudy} {et~al.}(2002){Rudy}, {Lynch}, {Mazuk}, {Venturini},
  {Puetter}, \& {H{\"o}flich}}]{Rudy_00cx}
{Rudy}, R.~J., {Lynch}, D.~K., {Mazuk}, S., {Venturini}, C.~C., {Puetter},
  R.~C., \& {H{\"o}flich}, P. 2002, ApJ, 565, 413

\bibitem[{{Stritzinger} {et~al.}(2006){Stritzinger}, {Leibundgut}, {Walch}, \&
  {Contardo}}]{Stritzinger}
{Stritzinger}, M., {Leibundgut}, B., {Walch}, S., \& {Contardo}, G. 2006, \aap,
  450, 241

\bibitem[{{Swartz}(1991)}]{Swartz_1991}
{Swartz}, D.~A. 1991, ApJ, 373, 604

\bibitem[{{Thielemann} {et~al.}(1986){Thielemann}, {Nomoto}, \&
  {Yokoi}}]{Thielemann_w7}
{Thielemann}, F.-K., {Nomoto}, K., \& {Yokoi}, K. 1986, A\&A, 158, 17

\bibitem[{{Timmes} {et~al.}(2003){Timmes}, {Brown}, \&
  {Truran}}]{Timmes_metallicity}
{Timmes}, F.~X., {Brown}, E.~F., \& {Truran}, J.~W. 2003, ApJ, 590, L83

\bibitem[{{Valentini} {et~al.}(2003){Valentini}, {Di Carlo}, {Massi}, {Dolci},
  {Arkharov}, {Larionov}, {Pastorello}, {Di Paola}, {Benetti}, {Cappellaro},
  {Turatto}, {Pedichini}, {D'Alessio}, {Caratti o Garatti}, {Li Causi},
  {Speziali}, {Danziger}, \& {Tornamb{\'e}}}]{Valentini_03E}
{Valentini}, G. {et~al.} 2003, ApJ, 595, 779

\bibitem[{{Wang} {et~al.}(2003){Wang}, {Baade}, {H{\" o}flich}, {Khokhlov},
  {Wheeler}, {Kasen}, {Nugent}, {Perlmutter}, {Fransson}, \&
  {Lundqvist}}]{Wang_01el}
{Wang}, L. {et~al.} 2003, ApJ, 591, 1110

\bibitem[{{Wheeler} {et~al.}(1998){Wheeler}, {Hoeflich}, {Harkness}, \&
  {Spyromilio}}]{Wheeler_IR}
{Wheeler}, J.~C., {Hoeflich}, P., {Harkness}, R.~P., \& {Spyromilio}, J. 1998,
  ApJ, 496, 908

\end{thebibliography}

\end{document}